\documentclass[12pt, draftclsnofoot, onecolumn]{IEEEtran}
\usepackage{cite}
\usepackage{amsmath}
\usepackage{amsmath,amssymb,amsfonts}
\usepackage{algorithm}
\usepackage{algorithmic}
\usepackage{subeqnarray}
\usepackage{cases}
\usepackage{enumitem}
\usepackage{booktabs}
\usepackage{color}
\usepackage{multirow}
\usepackage{tabularx}
\usepackage{textcomp}
\usepackage{xcolor}
\usepackage{graphicx} 
\usepackage{subfigure}
\usepackage{bbding} 
\usepackage{hyperref} 

\usepackage[normalem]{ulem}

\begin{document}

\title{Joint Communication and Sensing in RIS-enabled mmWave Networks\\}

\author{Lu Wang, Luis F. Abanto-Leon, and Arash Asadi


\thanks{Lu Wang, Luis F. Abanto-Leon, and Arash Asadi are with the Department of Computer Science, Technische Universität Darmstadt, Darmstadt 64289, Germany (e-mail: lwang@wise.tu-darmstadt.de; labanto@seemoo.tu-darmstadt.de; aasadi@wise.tu-darmstadt.de).}
}

\maketitle

\begin{abstract} 
Empowering cellular networks with augmented sensing capabilities is a key research area in sixth generation (6G) communication systems. Recently, we have witnessed a plethora of efforts to devise solutions that integrate sensing capabilities into communication systems, i.e., joint communication and sensing (JCAS). However, most prior works do not consider the impact of reconfigurable intelligent surfaces (RISs) on JCAS systems at millimeter-wave (mmWave) bands. Given that RISs are expected to become an integral part of future cellular systems, it is important to investigate their potential in cellular networks beyond communication goals. To comply with the waveform utilized in current cellular systems, this paper studies mmWave orthogonal frequency-division multiplexing (OFDM) JCAS systems in the presence of RISs. Specifically, we jointly design the hybrid beamforming and RIS phase shifts to guarantee the sensing functionalities via minimizing the beampattern mean square error at RIS, subject to signal-to-interference-plus-noise (SINR) and power constraints. The non-convexity of the investigated problem poses a challenge which we address by proposing a solution based on the penalty method and manifold-based alternating direction method of multipliers (ADMM). Simulation results demonstrate that both sensing and communication capabilities improve when the RIS is adequately designed. In addition, we discuss the tradeoff between sensing and communication.

\end{abstract}


\section{Introduction}
Joint communication and sensing (JCAS) is one of the core research areas in the sixth generation (6G) communication systems. By integrating sensing and communication into one physical unity, JCAS adds a new dimension to network intelligence allowing users and operators to simultaneously communicate and obtain/sense information from their surrounding environment~\cite{bFanoverview}. This opens up new possibilities for emerging use-cases involving autonomous or assisted navigation and smart home monitoring. Meanwhile, JCAS systems are expected to be deployed in the millimeter-wave (mmWave) bands which provide the benefit of larger bandwidth to achieve higher throughput and sensing resolution. However, integrating sensing and communication is a challenging task, in particular, at mmWave frequencies where signals are prone to high attenuation and blockage due to the short wavelength and high path loss~\cite{bSenCom1}. 

To tackle these problems, base stations (BSs) and users rely on highly directional beamforming, which requires efficient beamforming in JCAS scenarios not only towards the communication end-points but also the sensed targets. In mmWave JCAS systems, digital beamforming is difficult to realize due to the high cost of having one dedicated radio frequency (RF) chain for each antenna element, which motivates the exploitation of hybrid beamforming. Furthermore, to circumvent the blockages met at mmWave frequencies, reconfigurable intelligent surfaces (RISs) are a promising solution, which are expected to be deployed massively in 6G communication systems. RISs are planar electromagnetic surfaces consisting of a large number of controllable reflecting elements. Depending on the design, RISs can be controlled to alter the phase and amplitude of the incident signal without complex decoding, encoding, and radio-frequency processing operations. This makes RISs a scalable solution to circumvent the issue of susceptibility to blockages in a cost-effective manner. Therefore, we consider introducing the RIS into JCAS networks, which makes the system design of JCAS become even more challenging. Particularly, the beamforming design in such scenarios should simultaneously consider the mobile users for communication, the targets for sensing, and the phase configuration of RIS.

\subsection{Prior Works}
To date, the majority of existing works focused on either JCAS~\cite{bradarwaveform, bFLTWC2018, bFLwaveform, bPareto, bTradeoff1, bTradeoff2, bTradeoff3, bComSensing1, bComSensing2, bSenCom1, bSenCom2, bFLmmWave1, bZiyangC1, bZiyangC2, bZiyangC4, bZiyangC5, bFLmmWave2} or RISs~\cite{bRIS1, bRISSR, bQQW1, bRISEE, bNOMA, bMEC, bPenaltyM, bradar1, bradar2, bPeking1}. Hence, we first provide an overview of the state of the art on JCAS and RISs respectively, and then review the works that address JCAS in the presence of RISs.

{\bf JCAS}. 
The existing works in JCAS mostly focused on the waveform design~\cite{bradarwaveform, bFLTWC2018, bFLwaveform, bPareto}, performance tradeoff optimization~\cite{bTradeoff1, bTradeoff2, bTradeoff3}, communication-assisted sensing~\cite{bComSensing1, bComSensing2}, and sensing-assisted communication~\cite{bSenCom1, bSenCom2}. In recent years, we have observed more works on the design of JCAS at mmWave bands~\cite{bSenCom1, bSenCom2, bFLmmWave1, bZiyangC1, bZiyangC2, bZiyangC4, bZiyangC5, bFLmmWave2}. Specifically, in \cite{bFLmmWave1, bZiyangC1, bZiyangC2}, the authors designed hybrid beamformers by formulating a tradeoff problem between communication and sensing. Besides, the works in~\cite{bZiyangC4, bZiyangC5} designed the hybrid beamforming for multi-carrier JCAS systems. The works in~\cite{bSenCom1, bSenCom2, bFLmmWave2} investigated beam training, tracking, and prediction in JCAS systems at mmWave bands.

{\bf RIS}. Intelligent surfaces have been extensively studied from a communication perspective in recent years~\cite{bRIS1} for different design goals including sum-rate maximization~\cite{bRISSR}, power minimization~\cite{bQQW1}, and energy efficiency maximization~\cite{bRISEE}, and for various scenarios including non-orthogonal multiple access~\cite{bNOMA}, mobile edge computing~\cite{bMEC}, and simultaneous wireless information and power transfer~\cite{bPenaltyM}. A few works have also exploited RISs for sensing but mainly focused on detection and estimation aspects, such as improving target detection probability~\cite{bradar1}, target parameters estimation~\cite{bradar2}, and human gesture recognition~\cite{bPeking1}. All these foregoing applications of RIS in either communication systems or sensing systems demonstrate the potential of exploiting RIS in JCAS system.

{\bf RIS-assisted JCAS.} The number of works investigating JCAS in the presence of RISs~\cite{bRISsense1, bRISsense2, bRISsense3, bRISsense4, bWXY1, bWXY2, bRISCOM3, bRISCOM4, bRISCOM5, bSPAWC, b1T1UE, b1T1UE2, bTongWei, bRangLiu1, bSTARS, bRangLiu2, bXianxinSong2, bMengHua, bXianxinSongWCNC, bIndiaRS2, bIndiaRS3, bRISBPgap, bPekingHlZ, bLearning} is significantly less than the aforementioned two categories. The works~\cite{bRISsense1, bRISsense2, bRISsense3, bRISsense4, bWXY1, bWXY2, bRISCOM3, bRISCOM4, bRISCOM5} considered JCAS-enabled base stations (BSs) but the RIS usage was either limited to sensing/localization~\cite{bRISsense1, bRISsense2, bRISsense3, bRISsense4} or communication~\cite{bWXY1, bWXY2, bRISCOM3, bRISCOM4, bRISCOM5}. The authors of \cite{bSPAWC, b1T1UE, b1T1UE2} designed JCAS systems leveraging RISs for scenarios with a single user and a single target. Furthermore, the authors of~\cite{bTongWei, bRangLiu1, bSTARS, bRangLiu2, bXianxinSong2, bMengHua} investigated RIS-assisted JCAS systems for one single target and multiple users using different sensing metrics. To be specific, subject to signal-to-interference-plus-noise-ratio (SINR) or sum rate constraint for communication users, the sensing performance was optimized with different goals, namely radar SINR maximization in~\cite{bTongWei, bRangLiu1}, Cram{\'e}r-Rao lower bound minimization in~\cite{bSTARS, bRangLiu2}, beampattern gain towards targets maximization~\cite{bXianxinSong2, bMengHua}. Single-user and multi-target systems were studied in~\cite{bXianxinSongWCNC}, where the minimum beampattern gain for each of the targets was maximized subject to the signal-to-noise-ratio (SNR) requirement for one single user. As a step forward, multi-target and multi-user systems were investigated in~\cite{bIndiaRS2, bIndiaRS3, bRISBPgap, bPekingHlZ, bLearning}. Specifically, the authors of~\cite{bIndiaRS2, bIndiaRS3, bRISBPgap} maximized the minimum target illumination power or minimized the beampattern design mean square error (MSE) to guarantee the sensing performance while satisfying the SINR constraint for users. The work~\cite{bPekingHlZ} investigated the tradeoff between the sensing mutual information of targets and the SINR of users. The work~\cite{bLearning} jointly designed the beamforming and phase shifts in the terahertz band to maximize the sum rate of communication users by satisfying a beampattern MSE constraint.
\begin{table*}[!t]
	\caption{RIS-assisted JCAS Related Works.}
	\label{Tab2}
	\setlength{\tabcolsep}{8pt} 
	\centering
	\begin{tabular}{cccccccc}
		\toprule 
		\multirow{2}{*}{Ref.} & \multicolumn{2}{c}{System}  & \multirow{2}{*}{Spectrum} & \multirow{2}{*}{OFDM} & \multirow{2}{*}{Beamforming} & \multicolumn{2}{c}{RIS} \\
		\cline{2-3} \cline{7-8} & Users & Targets &    &    &    & Communication & Sensing \\
		\midrule 
		\cite{bRISsense1}& \XSolidBrush & Single & Sub-6 GHz & \XSolidBrush & \XSolidBrush & \XSolidBrush & \Checkmark \\
		\cite{bRISsense2}& \XSolidBrush & Single & Sub-6 GHz & \XSolidBrush & Digital beamforming & \XSolidBrush & \Checkmark \\
		\cite{bRISsense3, bRISsense4}& \XSolidBrush & Multiple & Sub-6 GHz & \XSolidBrush & Digital beamforming & \XSolidBrush & \Checkmark \\						
		\cite{bWXY1, bWXY2} &Multiple &Multiple & Sub-6 GHz & \XSolidBrush & \XSolidBrush & \Checkmark & \XSolidBrush \\
		\cite{bRISCOM3, bRISCOM4} &Multiple &Multiple & Sub-6 GHz & \XSolidBrush & Digital beamforming & \Checkmark & \XSolidBrush \\
        \cite{bRISCOM5} &Multiple &Multiple & mmWave & \XSolidBrush & Digital beamforming & \Checkmark &\XSolidBrush\\	
		\cite{bSPAWC} &Single &Single & mmWave & \XSolidBrush & \XSolidBrush & \Checkmark & \Checkmark \\		\cite{b1T1UE, b1T1UE2} & Single & Single & Sub-6 GHz & \XSolidBrush & Digital beamforming & \Checkmark & \Checkmark \\
		\cite{bTongWei} & Multiple & Single & Sub-6 GHz & \Checkmark & Digital beamforming & \Checkmark & \Checkmark \\	
		\cite{bRangLiu1, bSTARS, bRangLiu2, bXianxinSong2, bMengHua} & Multiple & Single & Sub-6 GHz & \XSolidBrush & Digital beamforming & \Checkmark & \Checkmark \\
        \cite{bXianxinSongWCNC} & Single & Multiple & Sub-6 GHz & \XSolidBrush & Digital beamforming & \Checkmark & \Checkmark \\        
		\cite{bIndiaRS2, bIndiaRS3, bPekingHlZ, bRISBPgap} & Multiple & Multiple & Sub-6 GHz & \XSolidBrush & Digital beamforming & \Checkmark & \Checkmark \\ 		
		\cite{bLearning} & Multiple & Multiple & Terahertz & \XSolidBrush & Digital beamforming & \Checkmark & \Checkmark \\
		This paper & Multiple & Multiple & mmWave & \Checkmark & Hybrid beamforming & \Checkmark & \Checkmark \\	
		\bottomrule 
	\end{tabular}
\end{table*}

However, the beamforming solutions in the aforementioned RIS-based JCAS networks only apply to fully-digital beamforming, which is difficult to realize at higher frequencies. Table~\ref{Tab2} summarizes the state of the art in RIS-assisted JCAS research. {\it We highlight that none of the aforementioned works has considered the RIS-assisted JCAS design in mmWave orthogonal frequency-division multiplexing (OFDM) systems, which requires the joint optimization of the RIS phase shifts and hybrid beamforming, while taking into account the multi-carrier nature of the modern wireless communication system.} Next, we discuss our motivation and contributions.

\subsection{Contributions}

In this article, we investigate RIS-assisted OFDM JCAS systems in mmWave cellular networks. Aligned with 5G new radio (NR) specifications, we adopt OFDM as the physical waveform of the system and consider multiple targets and users which experience different channel conditions on each sub-carrier~\cite{bSVchannel}. To improve the performance of communication and sensing, we jointly design the hybrid beamforming and RIS phase shifts, thus guaranteeing $(i)$ the communication for mobile users and $(ii)$ the sensing of targets. To realize the aforementioned goals, we formulate an optimization problem that proves to be non-convex, in particular, due to variable coupling, non-convex objective function and SINR constraints, and constant-modulus constraints that are inherent to the RIS elements and hybrid beamforming. Therefore, we propose a manifold-based alternating direction method of multipliers (ADMM) algorithm which is capable of coping with the complex structure of the formulated problem. In the following, we summarize our contributions in more detail.

\begin{itemize} 
	
	\item To the best of our knowledge, this is the first effort toward modeling an OFDM JCAS cellular system at mmWave bands in the presence of RISs. To guarantee the sensing performance, we minimize the difference between the designed and reference beampatterns at the RIS across all sub-carriers. In particular, the reference beampattern is designed to have main lobes in the directions of targets. As constraints, we incorporate SINR requirements for each user and each sub-carrier in order to guarantee the communication performance. In addition, we consider the transmit power constraint and constant-modulus phase shifts due to the hybrid beamforming and RIS.

	\item To solve the non-convex problem, we propose a scheme based on the penalty method and the ADMM algorithm. Specifically, we first introduce an auxiliary variable to decouple the RF beamformer and baseband beamformer. Then, we use the penalty method to overcome the non-convex SINR constraints by integrating them into the objective function. To tackle the newly transformed problem, we adopt a manifold-based ADMM framework, where each variable is optimized in an alternating and iterative manner until convergence.

	\item We evaluate our proposed scheme via extensive simulations under different parameter settings. Specifically, we assess its performance by analyzing the convergence, the sensing performance, and the communication performance. The sensing performance is evaluated via the beampattern MSE between the designed and reference beampatterns at the RIS, and the peak-side-lobe-ratio (PSLR) produced by the generated beampattern.
	The communication performance is evaluated via the feasibility ratio of SINR values above the predefined threshold. Furthermore, three benchmark schemes are included for comparison, namely fully-digital beamforming with RIS, fully-digital beamforming with random RIS phase shifts, and hybrid beamforming with random RIS phase shifts. The results show that our proposed scheme improves both sensing and communication performances with the assistance of RIS. The tradeoff between sensing and communication is also illustrated via comparing the SINR feasibility ratio and average SINR versus the beampattern MSE.
	
\end{itemize}

\subsection{Organization and Notation}
The remainder of this paper is organized as follows: Section \uppercase\expandafter{\romannumeral2} introduces the system model of the RIS-assisted mmWave OFDM JCAS system followed by the problem formulation. Section \uppercase\expandafter{\romannumeral3} presents the proposed solution and its computational complexity analysis. We provide simulation results and discussion in Section \uppercase\expandafter{\romannumeral5}. Finally, conclusions are summarized in Section \uppercase\expandafter{\romannumeral6}.

In this paper, bold-face uppercase and lowercase, ${\bf{A}}$ and ${\bf{a}}$, represent matrices and vectors, respectively. $\mathbb{E}\left\{  \cdot  \right\}$, ${\rm{tr}}\left(  \cdot  \right)$, and ${\mathcal O}\left(  \cdot  \right)$ stand for the expectation, the trace of a square matrix, and the computational complexity order. ${\left(  \cdot  \right)^{\rm{T}}}$ and ${\left(  \cdot  \right)^{\rm{H}}}$ denote the transpose and the Hermitian transpose of a matrix or vector. $\left|  \cdot  \right|$ and ${\left\| \cdot \right\|_{\rm{F}}}$ represent the absolute value of a scalar and the Frobenius norm of a matrix, respectively. The symbols $\otimes$, $\odot$, $\left\langle { \cdot, \cdot } \right\rangle $ represent the Kronecker product, the Hadamard product, and the Euclidean inner product, respectively. ${\nabla}$ denotes the Euclidean gradient. ${\rm{blkdiag}}\left(  \cdot  \right)$ and ${\rm{diag}}\left(  \cdot  \right)$ are functions to form a block diagonal or diagonal matrix with the elements in $\left(  \cdot  \right)$, respectively. ${\mathbb{C}}$ and ${\mathbb{R}}$ denote the complex and real numbers, respectively.

\section{System Model and Problem Formulation}
\begin{figure}[htbp]
	\centering
	\includegraphics[width=0.5\linewidth]{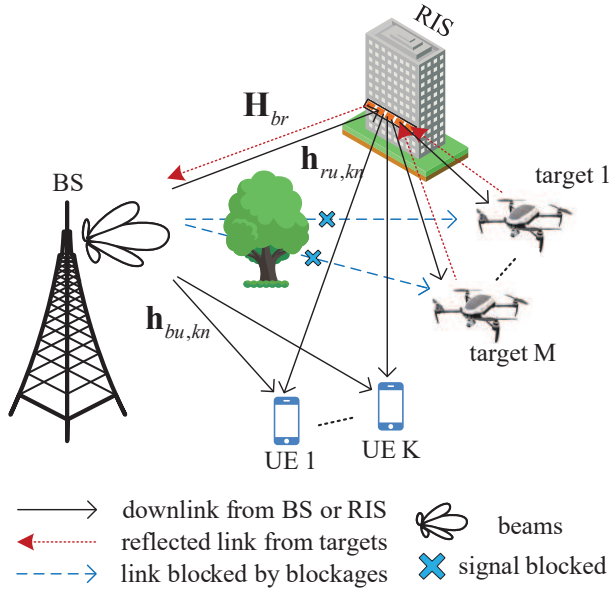}
	\caption{Overview of the RIS-assisted JCAS system at mmWave frequency.}
	\label{SysMod}
\end{figure}

We consider an OFDM JCAS system operating at mmWave frequencies, where an RIS is deployed to assist both communication and sensing, as shown in Fig.~\ref{SysMod}. The signals transmitted from the BS arrive at the user equipments (UEs) by direct transmission and indirect reflection via the RIS. For sensing, we consider the situation where the direct links between the BS and the targets are blocked. Thus, the RIS is deployed to circumvent the blockage and enable sensing.  The BS is equipped with ${{N}_{t}}$ antennas, which are spaced by half wavelength thus forming a uniform linear array (ULA). In addition, the BS has a small number of RF chains which is denoted by $N_{RF}$, such that $N_{RF} << {{N}_{t}} $. The BS serves $K$ single-antenna UEs and senses $M$ targets, simultaneously. The number of sub-carriers is denoted as $N_c$. The interval between adjacent sub-carriers is $\Delta f$. The RIS is configured with $R$ reflecting elements and modeled as a ULA structure\footnote{The problem formulation and solution regarding RIS in this paper can be applied to the RIS configured with uniform planer array (UPA), which is left for future work.}. Regarding the reflected path from the RIS, let ${{\theta }_{r}}\in [0,2\pi )$ denote the phase shift imposed by the $r$-th reflection element on the incident signals and $\Theta ={\rm{diag}}({{e}^{j{{\theta }_{1}}}},\cdots,{{e}^{j{{\theta }_{R}}}})$ group the phase shifts of all RIS elements\footnote{The amplitude of RIS is set as one to reduce the system design complexity and hardware cost.}. Due to the high path loss at mmWave band, it is assumed that the transmission signals reflected two or more times by the RIS are negligible and thus ignored~\cite{bQQW2}.

\subsection{Transmission Signals and Hybrid Beamforming}
Let ${\bf{X}} = {[{\bf{X}}_1^{\rm{T}},...,{\bf{X}}_n^{\rm{T}},...,{\bf{X}}_{{N_c}}^{\rm{T}}]^{\rm{T}}} \in {\mathbb{C}^{{N_c}K \times L}}$ denote the transmit symbols from the BS to all the $K$ UEs on all the $N_c$ sub-carriers, where ${{\bf{X}}_n} = {[{{\bf{x}}_{1,n}},...,{{\bf{x}}_{k,n}},...,{{\bf{x}}_{K,n}}]^{\rm{T}}} \in {\mathbb{C}^{K \times L}}$ stands for the transmit symbols for all the $K$ UEs on the sub-carrier $n \in {\mathcal{N}_c}$. Besides, $L$ represents the length of symbols. ${{\bf{x}}_{k,n}} \in {\mathbb{C}^{L \times 1}}$ refers to the data for the UE $k$ on the sub-carrier $n$, satisfying $\mathbb{E}{\{ {{\bf{x}}_{k,n}}{\bf{x}}_{k,n}^{\rm{H}}\} }={{\bf{I}}_L}$.

Hybrid beamforming consists of digital beamforming (a.k.a. baseband precoding) and analog beamforming (a.k.a. RF analog processing). The baseband digital beamforming is sub-carrier dependent whereas RF analog beamforming is identical for all the sub-carriers \cite{bRFBFBBBF}. The transmit symbols are first digitally processed by the precoding matrix ${{\bf{W}}^{BB}}$ in the frequency domain as
\begin{equation}\label{Xbf}
{{\bf{X}}^{BB}} = {{\bf{W}}^{BB}}{\bf{X}} = {[{\left( {{\bf{X}}_1^{BB}} \right)^{\rm{T}}},...,{\left( {{\bf{X}}_{{N_c}}^{BB}} \right)^{\rm{T}}}]^{\rm{T}}} \in {\mathbb{C}}{^{{N_c}{N_{RF}} \times L}},
\end{equation}
where ${{\bf{W}}^{BB}} = {\rm{blkdiag}}({\bf{W}}_1^{BB},...,{\bf{W}}_n^{BB},...,{\bf{W}}_{{N_c}}^{BB}) \in {\mathbb{C}^{{N_c}{N_{RF}} \times {N_c}K}}$ collects the precoding matrices for all the $K$ UEs on all the $N_c$ sub-carriers, ${\bf{W}}_n^{BB} = [{\bf{w}}_{1,n}^{BB},...,{\bf{w}}_{k,n}^{BB},...,{\bf{w}}_{K,n}^{BB}] \in {\mathbb{C}^{{N_{RF}} \times K}}$ represents the baseband precoding matrix for all the $K$ UEs on the sub-carrier $n$, and ${\bf{w}}_{k,n}^{BB} \in {\mathbb{C}^{{N_{RF}} \times 1}}$ is the baseband precoding vector for the UE $k$ on the sub-carrier $n$. Besides, ${\bf{X}}_n^{BB} = {\bf{W}}_n^{BB}{{\bf{X}}_n} \in {\mathbb{C}^{{N_{RF}} \times L}}$ is the precoded signal after digital beamforming for all the $K$ UEs on the sub-carrier $n$.

Then, the inverse discrete
Fourier transform (IDFT) operation is applied to ${\bf{X}}^{BB}$ thus yielding
\begin{equation}\label{Xidft}
{{{\bf{X}}^{ID}} = ({{\bf{F}}^{\rm{H}}} \otimes {{\bf{I}}_{{N_{RF}}}}){{\bf{W}}^{BB}}{\bf{X}} \in {\mathbb{C}^{{N_c}{N_{RF}} \times L}} = {{[{{\left( {{\bf{X}}_1^{ID}} \right)}^{\rm{T}}},...,{{\left( {{\bf{X}}_{{N_c}}^{ID}} \right)}^{\rm{T}}}]}^{\rm{T}}}} \in {\mathbb{C}}{^{{N_c}{N_{RF}} \times L}},
\end{equation}
where ${\bf{F}} \in {\mathbb{C}^{{N_{c}} \times {N_{c}}}}$ is the discrete
Fourier transform (DFT) matrix. The $(n,n')$-th element of ${\bf{F}}$ $(n\in {\mathcal{N}_c},n'\in {\mathcal{N}_c})$ is calculated as follows
\begin{equation}\label{Fele}
{F_{n,n'}} = \frac{1}{{\sqrt {{N_c}} }}{e^{ - \frac{{j2\pi }}{{{N_c}}}(n - 1)(n' - 1)}}.
\end{equation}
Therefore, $({{\bf{F}}^{\rm{H}}} \otimes {{\bf{I}}_{{N_{RF}}}})\in {\mathbb{C}^{N_c{N_{RF}} \times N_c{N_{RF}}}}$ in (\ref{Xidft}) corresponds to a $N_c$-point IDFT for all the $N_{RF}$ RF chains. In addition, ${\bf{X}}_n^{ID} = ({\bf{f}}_n^{\rm{H}} \otimes {{\bf{I}}_{{N_{RF}}}}){\bf{W}}^{BB}{\bf{X}} \in {\mathbb{C}^{{N_{RF}} \times L}}$ in (\ref{Xidft}) is the transmitted signal after the IDFT operation for all the $N_{RF}$ RF chains on the sub-carrier $n$, where ${{\bf{f}}_n}\in {\mathbb{C}^{{N_c} \times 1}}$ is the $n$-th column of ${\bf{F}}$, corresponding to the DFT operation on the sub-carrier $n$.

Within the OFDM structure, the cyclic prefix is then added at the beginning of each OFDM symbol to deal with inter-symbol interference. Afterwards, the signals are up-converted, and RF analog beamforming is performed with the RF precoder ${{\bf{W}}^{RF}} = {\rm{blkdiag}}({\bf{W}}_1^{RF},...,{\bf{W}}_{{N_c}}^{RF}) \in {\mathbb{C}^{N_c N_t \times N_c N_{RF}}}$ in the time domain. Since the RF precoder is the same for all sub-carriers, we use ${{{{\bf{\tilde W}}}^{RF}}}$ to denote the RF precoder on each sub-carrier, such that ${\bf{W}}_1^{RF} = {\bf{W}}_2^{RF} =  \cdots  = {\bf{W}}_{{N_c}}^{RF} = {{{\bf{\tilde W}}}^{RF}} \in {\mathbb{C}^{{N_t} \times {N_{RF}}}}$. Therefore, the transmitted signal of all the $K$ UEs on all the $N_c$ sub-carriers after RF analog beamforming can be expressed as follows
\begin{equation}\label{XRF}
{{\bf{X}}^{RF}} = {{\bf{W}}^{RF}}({{\bf{F}}^{\rm{H}}} \otimes {{\bf{I}}_{{N_{RF}}}}){{\bf{W}}^{BB}}{\bf{X}} \in {\mathbb{C}^{{N_c}{N_t} \times L}} = {[{\left( {{\bf{X}}_1^{RF}} \right)^{\rm{T}}},...,{\left( {{\bf{X}}_{{N_c}}^{RF}} \right)^{\rm{T}}}]^{\rm{T}}} \in {\mathbb{C}}{^{{N_c}{N_t} \times L}},
\end{equation}
where ${\bf{X}}_n^{RF} = {{{\bf{\tilde W}}}^{RF}}({\bf{f}}_n^{\rm{H}} \otimes {{\bf{I}}_{{N_{RF}}}}){\bf{W}}^{BB}{\bf{X}}\in {\mathbb{C}^{N_t \times L}}$ is the signal for all the $K$ UEs on the $n$-th sub-carrier after analog beamforming.

\subsection{Communication and Channel Model}
\label{SecChannel}

The signal ${\bf{X}}^{RF}$ is transmitted from the BS to $K$ UEs through the direct link and the indirect link reflected by the RIS. The received signals can be expressed as follows
\begin{equation}\label{Yc}
{{\bf{Y}}_c} = ({\bf{H}}_{bu}^{\rm{H}} + {\bf{H}}_{ru}^{\rm{H}}\hat \Theta {\bf{H}}_{br}^{\rm{H}}){{\bf{W}}^{RF}}({{\bf{F}}^{\rm{H}}} \otimes {{\bf{I}}_{{N_{RF}}}}){{\bf{W}}^{BB}}{\bf{X}} + {{\bf{N}}_c},
\end{equation}
where ${{\bf{N}}_c}$ is the complex Gaussian noise during the transmission with zero mean and variance $\sigma _c^2$. $\hat \Theta = {\rm{blkdiag}}(\Theta ,...,\Theta )\in {{\mathbb{C}}^{{{N}_{c}R}\times {N}_{c}R}}$ is $N_c$ repetitions of $\Theta$ to match the matrix multiplication. ${{\bf{H}}_{bu}} = {\rm{blkdiag}}({{\bf{H}}_{bu,1}},...,{{\bf{H}}_{bu,n}},...,{{\bf{H}}_{bu,{N_c}}})\in {{\mathbb{C}}^{{{N}_{c}{N}_{t}}\times {N}_{c}K}}$, ${{\bf{H}}_{br}} = {\rm{blkdiag}}({{\bf{H}}_{br,1}},...,{{\bf{H}}_{br,n}},...,$ ${{\bf{H}}_{br,{N_c}}})\in {{\mathbb{C}}^{{{N}_{c}{N}_{t}}\times {{N}_{c}R}}}$ and ${{\bf{H}}_{ru}} = {\rm{blkdiag}}({{\bf{H}}_{ru,1}},...,{{\bf{H}}_{ru,n}},...,{{\bf{H}}_{ru,{N_c}}}) \in {{\mathbb{C}}^{{{N}_{c}R}\times {N}_{c}K}}$ are the channels on all the $N_c$ sub-carriers from the BS to $K$ UEs, from the BS to the RIS, and from the RIS to $K$ UEs, respectively.

For each sub-carrier $n$, ${{\bf{H}}_{bu,n}} = [{{\bf{h}}_{bu,1n}},...,{{\bf{h}}_{bu,kn}},...,{{\bf{h}}_{bu,Kn}}]\in {{\mathbb{C}}^{{{N}_{t}}\times K}}$ is the channel from the BS to $K$ UEs, where ${{\bf{h}}_{bu,kn}}\in {{\mathbb{C}}^{{{N}_{t}}\times 1}}$ is the channel from the BS to the UE $k$ on the sub-carrier $n$, modelled as the extended Saleh-Valenzuela model~\cite{bSVchannel} given by
\begin{equation}\label{Hbkn}
{{\bf{h}}_{bu,kn}} = \sqrt {\frac{{{N_t}}}{{{N_{cl}}{N_p}}}} \sum\limits_c^{{N_{cl}}} {\sum\limits_p^{{N_p}} {\alpha _{c,p}^{bu}} {\bf{a}}_b^{\rm{H}}({\phi _{b,c,p}}){e^{ - j\frac{{2\pi {\psi _c}n}}{{{N_c}}}}}},
\end{equation}
where ${N_{cl}}$ and ${N_p}$ represent the number of clusters and scattering paths in each cluster. ${\psi _c}$ is the phase shift of $c$-th cluster. For the $p$-th scattering path of the $c$-th cluster, ${\alpha _{c,p}^{bu}}$ and ${\phi _{b,c,p}}$ are the complex gain and the angle of departure (AoD), and ${{\bf{a}}_b}({\phi _{b,c,p}})$ is the corresponding transmit array response vector. Given the ULA structure with $N_t$ antennas at the BS, the transmit array response vector is expressed as
\begin{equation}\label{ARV}
{{\bf{a}}_{b}}(\phi ) = \sqrt {\frac{1}{{{N_t}}}} {[1,{e^{j\frac{{2\pi }}{\lambda }d\sin (\phi )}},...,{e^{j\frac{{2\pi }}{\lambda }d({N_t} - 1)\sin (\phi )}}]^{\rm{T}}},
\end{equation}
where $d$ and $\lambda$ are the antenna space and wavelength respectively, satisfying the relation $d = {\lambda  \mathord{\left/
		{\vphantom {\lambda  2}} \right.
		\kern-\nulldelimiterspace} 2}$.
Similarly, the channel ${{\bf{H}}_{br,n}} \in {{\mathbb{C}}^{{{N}_{t}}\times R}}$ from the BS to the RIS on the sub-carrier $n$ is given by
\begin{equation}\label{Hbrn}
{{\bf{H}}_{br,n}} = \sqrt {\frac{{{N_t}R}}{{{N_{cl}}{N_p}}}} \sum\limits_c^{{N_{cl}}} {\sum\limits_p^{{N_p}} {\alpha _{c,p}^{br}} {{\bf{a}}_r}({\phi _{r,c,p}}){\bf{a}}_b^{\rm{H}}({\phi _{b,c,p}}){e^{ - j\frac{{2\pi {\psi _c}n}}{{{N_c}}}}}},
\end{equation}
where for the $p$-th scattering path of the $c$-th cluster, ${{\phi _{r,c,p}}}$ is the angle of arrival (AoA) at RIS, ${\alpha _{c,p}^{br}}$ is the complex gain, ${{{\bf{a}}_r}({\phi _{r,c,p}})}$ is the receive antenna array response at RIS with the similar expression as (\ref{ARV}) but replacing the number of antennas with the number of RIS elements $R$.
Likewise, ${{\bf{H}}_{ru}} = [{{\bf{h}}_{ru,1n}},...,{{\bf{h}}_{ru,kn}},...,{{\bf{h}}_{bu,Kn}}]\in {{\mathbb{C}}^{{R}\times K}}$ is the channel from the RIS to $K$ UEs on the sub-carrier $n$. ${{\bf{h}}_{ru,kn}}\in {{\mathbb{C}}^{{R}\times 1}}$ is the channel from the RIS to the UE $k$ on the sub-carrier $n$, given by
\begin{equation}\label{Hrkn}
{{\bf{h}}_{ru,kn}} = \sqrt {\frac{R}{{{N_{cl}}{N_p}}}} \sum\limits_c^{{N_{cl}}} {\sum\limits_p^{{N_p}} {\alpha _{c,p}^{ru}} {\bf{a}}_b^{\rm{H}}({\phi _{r,c,p}}){e^{ - j\frac{{2\pi {\psi _c}n}}{{{N_c}}}}}},
\end{equation}
where ${\alpha _{c,p}^{ru}}$ is the complex gain of the $p$-th scattering path in the $c$-th cluster between RIS and UE $k$.

We assume that the channel state information (CSI) of all channels is perfectly estimated, which can be realized by the current channel estimation techniques~\cite{bCSI1}. By denoting $({{\bf{H}}_{bu}} + {{\bf{H}}_{br}}{{\hat \Theta }^{\rm{H}}}{{\bf{H}}_{ru}})$ as ${\bf{\tilde H}}$, (\ref{Yc}) is recast as
\begin{equation}\label{Ycc}
{\bf{Y}}_c = {{{\bf{\tilde H}}}^{\rm{H}}}{{\bf{W}}^{RF}}({{\bf{F}}^{\rm{H}}} \otimes {{\bf{I}}_{{N_{RF}}}}){{\bf{W}}^{BB}}{\bf{X}} + {{\bf{N}}_c}.
\end{equation}
To reflect the communication performance, SINR is utilized as the communication metric in this paper. Accordingly, the SINR of UE $k$ on the sub-carrier $n$ is given by
\begin{equation}
{\gamma _{k,n}} = \frac{{{{\left| {{{{\bf{\tilde h}}}_{k,n}}{{{\bf{\tilde W}}}^{RF}}({\bf{f}}_n^{\rm{H}} \otimes {{\bf{I}}_{{N_{RF}}}}){\bf{W}}_k^{BB}} \right|}^2}}}{{\sum\limits_{i = 1,i \ne k}^K {{{\left| {{{{\bf{\tilde h}}}_{k,n}}{{{\bf{\tilde W}}}^{RF}}({\bf{f}}_n^{\rm{H}} \otimes {{\bf{I}}_{{N_{RF}}}}){\bf{W}}_i^{BB}} \right|}^2}}  + \sigma _c^2}}.
\end{equation}
where ${{{\bf{\tilde h}}}_{k,n}}$ is the corresponding combined channel for the UE $k$ on the sub-carrier $n$.

\subsection{Sensing Model}
The BS not only communicates with UEs, but also senses the environment by the signal reflection toward the targets. Due to the blockage between the BS and the targets, sensing is realized via the reflected signals from the RIS. To measure the sensing performance, we leverage the beampattern formed at the RIS towards the angles of all the targets. The beampattern towards the detection angle $\psi$ generated at the RIS is defined as follows\cite{bRISBPgap}
\begin{equation}\label{RISBP}
{\rm{B}}{{\rm{P}}_r}\left( \psi  \right) = \left\| {{{\bf{A}}^{\rm{H}}}\left( \psi  \right)\hat \Theta {\bf{H}}_{br}^{\rm{H}}{{\bf{W}}^{RF}}({{\bf{F}}^{\rm{H}}} \otimes {{\bf{I}}_{{N_{RF}}}}){{\bf{W}}^{BB}}} \right\|_{\rm{F}}^2,
\end{equation}
where ${\bf{A}} = {\rm{diag}}\left( {{{{\bf{\hat a}}}_1}\left( \psi  \right),...,{{{\bf{\hat a}}}_n}\left( \psi  \right),...,{{{\bf{\hat a}}}_{{N_c}}}\left( \psi  \right)} \right) \in {{\mathbb{C}}^{{{N}_{c}R}\times {{N}_{c}}}}$ is the array response vector at the RIS on the $N_c$ sub-carriers. ${{{\bf{\hat a}}}_n}\left( \psi  \right) = {[1,{e^{j2\pi d\sin (\psi )}},...,{e^{j2\pi d(R - 1)\sin (\psi )}}]^{\rm{T}}}\in {{\mathbb{C}}^{R\times 1}}$ is the array response vector at the RIS on the sub-carrier $n$.

\subsection{Problem Formulation}
In this paper, we aim to simultaneously optimize the performances of sensing and communication in the presence of RIS. To guarantee the communication performance, SINR is leveraged as the metric and the SINR of each UE on each sub-carrier is constrained to be not smaller than a predefined threshold. When the RIS is deployed to improve the sensing performance as shown in Fig.~\ref{SysMod}, the beampattern gain at the RIS is utilized as the sensing metric to reflect the sensing performance. We optimize the beampattern gain at the RIS by matching it with a reference beampattern while simultaneously satisfying the transmit power and SINR requirements by jointly designing the hybrid beamforming and RIS phase shifts. This aforementioned optimization problem is mathematically formulated as
\begin{subequations}\label{OptP1}
	\begin{align}
		({\rm{P}}1)\mathop {\min }\limits_{{{\bf{W}}^{RF}},{{\bf{W}}^{BB}},\Theta } & \sum\limits_\psi ^{{N_\psi }} {{{\left| {{\rm{B}}{{\rm{P}}_{\rm{r}}}\left( \psi  \right) - {{\rm{BP}}_{{\rm{ref}}}}\left( \psi  \right)} \right|}^2}}  \label{P1Obj}\\
		\mbox{s.t.}\quad
		&{\gamma _{k,n}} \ge {\Gamma _{k,n}},\forall k \in \mathcal{K},n \in \mathcal{N}_c, \label{CQoS}\\
		&\left\| {{{\bf{W}}^{RF}}\left( {{{\bf{F}}^{\rm{H}}} \otimes {{\bf{I}}_{{N_{RF}}}}} \right){{\bf{W}}^{BB}}} \right\|_{\rm{F}}^2 = {P_{\max }}, \label{CHB}\\
		& \left| {{{{\bf{\tilde W}}}^{RF}}(i,j)} \right| = 1,\forall i \in {\mathcal{N}_t},j \in {\mathcal{N}_{RF}}, \label{CRFB}\\
		&{\theta _r} \in [0,2\pi ),\forall r \in \mathcal{R}, \label{CRIS}
	\end{align}
\end{subequations}
where ${N_\psi }$ is the number of angles for all the interesting targets coving the detection range $[-{\pi  \mathord{\left/{\vphantom {\pi  {2,{\pi  \mathord{\left/{\vphantom {\pi  2}} \right.			\kern-\nulldelimiterspace} 2}}}} \right.\kern-\nulldelimiterspace} {2,{\pi  \mathord{\left/	{\vphantom {\pi  2}} \right.\kern-\nulldelimiterspace} 2}}}]$. ${{\rm{BP}}_{{\rm{ref}}}}$ is the reference beampattern at RIS for all the sub-carriers. For each sub-carrier, the reference beampattern at RIS is obtained by performing the least square method used in~\cite{bFLTWC2018} between the reference beampattern to be solved and the ideal beampattern with all one values at the angles of targets. The constraint (\ref{CQoS}) guarantees the communication performance by ensuring that the SINR of UE $k$ on the sub-carrier $n$ is not smaller than a predefined minimum SINR requirement $\Gamma_{k,n}$. The constraints (\ref{CHB}) and (\ref{CRFB}) are the power allocation for hybrid beamforming over all the sub-carriers, which is supposed to consume all the provided power by BS for better sensing effects~\cite{bFLTWC2018}, making the problem ${\rm P}1$ more difficult to solve. Assuming the reflection amplitudes of RIS elements are all one, the constraint (\ref{CRIS}) is the design of RIS phase shifts. Next, we design an ADMM algorithm based on the penalty method and manifold optimization to tackle the formulated problem.

\section{Proposed Solution}
In this section, we design the solution to the formulated problem ${\rm P}1$. From the objective function (\ref{P1Obj}) and the constraints (\ref{CQoS}) and (\ref{CHB}), we can observe that all the variables are coupled with each other, which makes the non-convex problem ${\rm P}1$ more challenging to solve. Considering the difficulty of directly solving the original problem ${\rm P}1$, we first transform problem ${\rm P}1$ into a more tractable form (i.e., problem ${\rm P}2$). Then, a manifold-based ADMM algorithm is proposed to obtain the solution. Details are described in the following.

To make the formulated problem more tractable, we first introduce an auxiliary variable ${\bf{W}} = {{\bf{W}}^{RF}}\left( {{{\bf{F}}^{\rm{H}}} \otimes {{\bf{I}}_{{N_{RF}}}}} \right){{\bf{W}}^{BB}} \in {\mathbb{C}^{{N_c}{N_t} \times {N_c}K}}$ to decouple the RF analog beamformer ${{\bf{W}}^{RF}}$ and baseband digital beamformer ${{\bf{W}}^{BB}}$. Then the problem ${\rm P}1$ becomes the following problem ${\rm P}2$.
\begin{subequations}\label{Opt3}
	\begin{align}
		({\rm P}2) \mathop {\min }\limits_{{{{{\bf{W}}^{RF}}}},{{{{\bf{W}}^{BB}}}},{\bf{W}},{\Theta}} &\sum\limits_\psi ^{{N_\psi }} {{{\left| {{{\bf{A}}^{\rm{H}}}\left( \psi  \right)\hat \Theta {\bf{H}}_{br}^{\rm{H}}{\bf{W}}{{\bf{W}}^{\rm{H}}}{{\bf{H}}_{br}}{{\hat \Theta }^{\rm{H}}}{\bf{A}}\left( \psi  \right) - {{\rm{BP}}_{{\rm{ref}}}}\left( \psi  \right)} \right|}^2}}  \\
		\mbox{s.t.}\quad
		&{\bf{W}} = {{\bf{W}}^{RF}}\left( {{{\bf{F}}^{\rm{H}}} \otimes {{\bf{I}}_{{N_{RF}}}}} \right){{\bf{W}}^{BB}}, \label{CEqu}\\
		&\frac{{{{\left| {{{{\bf{\tilde h}}}_{k,n}}{{\bf{w}}_{k,n}}} \right|}^2}}}{{\sum\limits_{i = 1,i \ne k}^K {{{\left| {{{{\bf{\tilde h}}}_{k,n}}{{\bf{w}}_{i,n}}} \right|}^2}}  + \sigma _c^2}} \ge {\Gamma _{k,n}},\forall k \in \mathcal{K},n \in \mathcal{N}_c, \label{CSINRW}\\
		&\left\| {\bf{W}} \right\|_{\rm{F}}^2 = {P_{\max }}, \label{CPwrW}\\
		&\left| {{{{\bf{\tilde W}}}^{RF}}(i,j)} \right| = 1,\forall i \in {\mathcal{N}_t},j \in {\mathcal{N}_{RF}}, \\
		&{\theta _r} \in [0,2\pi ),\forall r \in \mathcal{R}.
	\end{align}
\end{subequations}
Leveraging the equality constraint (\ref{CEqu}), we apply the ADMM framework to tackle the problem ${\rm P}2$. The augmented Lagrangian function of the problem ${\rm P}2$ is calculated as
\begin{equation}\label{ALag}
\begin{array}{l}
	L = \sum\limits_\psi ^{{N_\psi }} {{{\left| {{{\bf{A}}^{\rm{H}}}\left( \psi  \right)\hat \Theta {\bf{H}}_{br}^{\rm{H}}{\bf{W}}{{\bf{W}}^{\rm{H}}}{{\bf{H}}_{br}}{{\hat \Theta }^{\rm{H}}}{\bf{A}}\left( \psi  \right) - {\rm{B}}{{\rm{P}}_{{\rm{ref}}}}\left( \psi  \right)} \right|}^2}} \\
	+ \frac{{{\rho _1}}}{2}\left\| {{\bf{W}} + \frac{{\bf{\Lambda }}}{{{\rho _1}}} - {{\bf{W}}^{RF}}\left( {{{\bf{F}}^{\rm{H}}} \otimes {{\bf{I}}_{{N_{RF}}}}} \right){{\bf{W}}^{BB}}} \right\|_{\rm{F}}^2
\end{array}
\end{equation}
where ${\bf{\Lambda }}\in {{\mathbb{C}}^{{{N}_{c}{N}_t}\times {N}_{c}K}}$ and ${{\rho _1} > 0}$ are the Lagrangian multipliers and the penalty parameter, respectively. Then, under the ADMM framework, all the variables and Lagrangian multipliers $\{ {\bf{W}},{{\bf{W}}^{RF}},{{\bf{W}}^{BB}},{\Theta},{\bf{\Lambda }}\}$ update as follows:

\subsection{Update of Auxiliary Variable}
Given $\{ {{\bf{W}}^{RF}},{{\bf{W}}^{BB}},{\Theta},{\bf{\Lambda }}\}$, the sub-problem for the auxiliary variable ${\bf{W}}$ is expressed as
\begin{subequations}\label{OptAuxiliary}
	\begin{align}
		{{\bf{W}}^{(t)}} =& \arg \mathop {\min }\limits_{\bf{W}} {L}({\bf{W}},{{{\bf{W}}^{RF}}^{(t - 1)}},{{{\bf{W}}^{BB}}^{(t - 1)}},{{\Theta}^{(t - 1)}},{{\bf{\Lambda }}^{(t - 1)}}) \\
		\mbox{s.t.}\quad
		&\frac{{{{\left| {{{{\bf{\tilde h}}}_{k,n}}{{\bf{w}}_{k,n}}} \right|}^2}}}{{\sum\limits_{i = 1,i \ne k}^K {{{\left| {{{{\bf{\tilde h}}}_{k,n}}{{\bf{w}}_{i,n}}} \right|}^2}}  + \sigma _c^2}} \ge {\Gamma _{k,n}},\forall k \in \mathcal{K},n \in \mathcal{N}_c, \label{CSINRW1}\\
		&\left\| {\bf{W}} \right\|_{\rm{F}}^2 = {P_{\max }}.
	\end{align}
\end{subequations}
The existence of the SINR constraint (\ref{CSINRW1}) makes the problem (\ref{OptAuxiliary}) still challenging to solve. Therefore, the SINR inequality constraint is further integrated into the objective function by using the penalty method~\cite{bPenaltyM}. This means if the obtained solution satisfies the SINR constraint, it is equivalent to the original problem (\ref{OptAuxiliary}). Otherwise, the penalty is given to the unsuitable solution. Denoting $\left( {{{\bf{A}}^{\rm{H}}}\left( \psi  \right)\hat \Theta {\bf{H}}_{br}^{\rm{H}}} \right)$ further as ${{\bf{X}}_\psi }$, the new problem with the SINR penalty is expressed as follows
\begin{subequations}\label{OptSINRPenalty}
\begin{align}
{({\rm{P}}2.1)}&\begin{array}{*{20}{l}}
	{\mathop {\min }\limits_{\bf{W}} \left\{ {\sum\limits_\psi ^{{N_\psi }} {{{\left| {{{\bf{X}}_\psi }{\bf{W}}{{\bf{W}}^{\rm{H}}}{\bf{X}}_\psi ^{\rm{H}} - {\rm{B}}{{\rm{P}}_{{\rm{ref}}}}\left( \psi  \right)} \right|}^2}} } \right.}\\
	{\left. { + \frac{{{\rho _1}}}{2}\left\| {{\bf{W}} + \frac{{\bf{\Lambda }}}{{{\rho _1}}} - {{\bf{W}}^{RF}}\left( {{{\bf{F}}^{\rm{H}}} \otimes {{\bf{I}}_{{N_{RF}}}}} \right){{\bf{W}}^{BB}}} \right\|_{\rm{F}}^2 + {\rho _2}\sum\limits_{k = 1}^K {\sum\limits_{n = 1}^{{N_c}} {P_{k,n}^{\bf{W}}} } } \right\}}
\end{array} \label{P2.1Obj}\\
{{\rm{s.t.}}\quad }&{\left\| {\bf{W}} \right\|_{\rm{F}}^2 = {P_{\max }}.}\label{ConsW}
\end{align}
\end{subequations}
where $P_{k,n}^{\bf{W}} = {\left( {\min \left\{ {{\gamma _{k,n}} - {\Gamma _{k,n}},0} \right\}} \right)^2}$, and ${\rho _2}$ is the penalty parameter for SINR constraint, which usually updates from a small value to a large value.

Denoting ${{\bf{B}}_{k,n}} = {{{\bf{\tilde h}}}_{k,n}}{\left( {{{{\bf{\tilde h}}}_{k,n}}} \right)^{\rm{H}}}$ and ${{\bf{F}}_{k,n}} = {{\bf{w}}_{k,n}}{\left( {{{\bf{w}}_{k,n}}} \right)^{\rm{H}}}$, we have the following transformation
\begin{equation}\label{CSINRpenalty}
\sum\limits_{k = 1}^K {\sum\limits_{n = 1}^{Nc} {{{\left( {{\gamma _{k,n}} - {\Gamma _{k,n}}} \right)}^2} = \sum\limits_{k = 1}^K {\sum\limits_{n = 1}^{{N_c}} {{{\left( {\frac{{{\rm{tr}}\left( {{{\bf{B}}_{k,n}}{{\bf{F}}_{k,n}}} \right)}}{{{\rm{tr}}\left( {{{\bf{B}}_{k,n}}\sum\limits_{i = 1,i \ne k}^K {{{\bf{F}}_{i,n}}} } \right) + \sigma _c^2}} - {\Gamma _{k,n}}} \right)}^2}} } } }.
\end{equation}
From the perspective of optimization, (\ref{CSINRpenalty}) is equivalent to
\begin{equation}\label{CSINRpenalty1}
\sum\limits_{k = 1}^K {\sum\limits_{n = 1}^{{N_c}} {{{\left( {{\gamma _{k,n}} - {\Gamma _{k,n}}} \right)}^2}} }  = \sum\limits_{k = 1}^K {\sum\limits_{n = 1}^{{N_c}} {{{\left( {{{\gamma '}_{k,n}} - \sigma _c^2{\Gamma _{k,n}}} \right)}^2}} }, 
\end{equation}
where ${{\gamma '}_{k,n}}$ is given by
\begin{equation}\label{gamma'k}
{{\gamma '}_{k,n}} = \left( {1 + {\Gamma _{k,n}}} \right){\rm{tr}}\left( {{{\bf{B}}_{k,n}}{{\bf{F}}_{k,n}}} \right) - {\Gamma _{k,n}}{\rm{tr}}\left( {{{\bf{B}}_{k,n}}\sum\limits_{i = 1}^K {{{\bf{F}}_{i,n}}} } \right).
\end{equation}
Therefore, $P_{k,n}^{\bf{W}}$ is equivalent to
\begin{equation}\label{P_W^k}
P_{k,n}^{\bf{W}} = {\left( {\min \left\{ {{{\gamma '}_{k,n}} - \sigma _c^2{\Gamma _{k,n}},0} \right\}} \right)^2}.
\end{equation}

After integrating the SINR constraint into the objective function, only one constraint for power (\ref{ConsW}) is left which is equivalent to ${\left\| {\bf{W}} \right\|_{\rm{F}}} = \sqrt {{P_{\max }}}$. The problem ${\rm P}2.1$ can then be regarded as a non-constraint problem over manifold, where the feasible set is a ${{{N}_{c}{N}_t}\times {N}_{c}K}-1$ dimension complex hypersphere. Therefore, the problem ${\rm P}2.1$ can be rewritten as follows
\begin{equation}\label{OptAuxTrans}
\begin{array}{l}
	\mathop {\min }\limits_{{\bf{W}} \in {S_{\bf{W}}}} {f_{\bf{W}}} = \left\{ {\sum\limits_\psi ^{{N_\psi }} {{{\left| {{{\bf{X}}_\psi }{\bf{W}}{{\bf{W}}^{\rm{H}}}{\bf{X}}_\psi ^{\rm{H}} - {\rm{B}}{{\rm{P}}_{{\rm{ref}}}}\left( \psi  \right)} \right|}^2}} } \right.\\
	\left. { + \frac{{{\rho _1}}}{2}\left\| {{\bf{W}} + \frac{{\bf{\Lambda }}}{{{\rho _1}}} - {{\bf{W}}^{RF}}\left( {{{\bf{F}}^{\rm{H}}} \otimes {{\bf{I}}_{{N_{RF}}}}} \right){{\bf{W}}^{BB}}} \right\|_{\rm{F}}^2 + {\rho _2}\sum\limits_{k = 1}^K {\sum\limits_{n = 1}^{{N_c}} {P_{k,n}^{\bf{W}}} } } \right\},
\end{array}
\end{equation}
where $\mathcal{S}_{\bf{W}} = \left\{ {{\bf{W}}\in {{\mathbb{C}}^{{{N}_{c}{N}_t}\times {N}_{c}K}}{\left| {{{\left\| {\bf{W}} \right\|}_{\rm{F}}} = \sqrt {{P_{\max }}} } \right.} } \right\}$ represents the complex hypersphere manifold with the radius of $\sqrt {{P_{\max }}}$. To efficiently solve the problem (\ref{OptAuxTrans}), we leverage a Riemannian Conjugate Gradient (RCG)-based algorithm, which can realize the near-optimal solution with low complexity~\cite{bManifold1,bManifold2}.

Firstly, we define the tangent space over the manifold $\mathcal{S_{\bf{W}}}$ as ${T_{\bf{W}}}\mathcal{S_{\bf{W}}}$, which is the space consisting of all the tangent vectors over $\mathcal{S_{\bf{W}}}$, namely
\begin{equation}\label{TangentS}
{T_{\bf{W}}}{{\cal S}_{\bf{W}}} = \left\{ {{\bf{T}} \in {{\mathbb{C}}^{{N_c}{N_t} \times {N_c}K}}\left| {{\rm{Re}}\left\{ {{\bf{T}} \odot {{\bf{W}}^ * }} \right\}} \right. = \textbf{0}} \right\}.
\end{equation}
Then the Riemannian gradient of $\mathcal{S_{\bf{W}}}$ at point ${\bf{W}}$ is ${\rm{grad}}\left( {f_{\bf{W}}} \right)$, which is obtained by projecting the Euclidean gradient of ${f_{\bf{W}}}$ on the tangent space ${T_{\bf{W}}}\mathcal{S_{\bf{W}}}$, namely
\begin{equation}\label{RiemannianG}
{\rm{grad}}\left( {{f_{\bf{W}}}} \right) = {\rm{Pro}}{{\rm{j}}_{\bf{W}}}\left( {{f_{\bf{W}}}} \right) = \nabla {f_{\bf{W}}} - {\rm{Re}}\left\{ {\nabla {f_{\bf{W}}} \odot {{\bf{W}}^*}} \right\} \odot {\bf{W}}.
\end{equation}
The Euclidean gradient ${\nabla {{f_{\bf{W}}}}}$ of ${{f_{\bf{W}}}}$ is derived as
\begin{equation}\label{EucGradW}
\begin{array}{l}
	\nabla {f_{\bf{W}}} = 4\sum\limits_\psi ^{{N_\psi }} {\left( {{\bf{X}}_\psi ^{\rm{H}}{{\bf{X}}_\psi }{\bf{W}}{{\bf{W}}^{\rm{H}}}{\bf{X}}_\psi ^{\rm{H}}{{\bf{X}}_\psi }{\bf{W}} - {\bf{X}}_\psi ^{\rm{H}}{\rm{B}}{{\rm{P}}_{{\rm{ref}}}}\left( \psi  \right){{\bf{X}}_\psi }{\bf{W}}} \right)} \\
	+ {\rho _1}\left( {{\bf{W}} + \frac{{\bf{\Lambda }}}{{{\rho _1}}} - {{\bf{W}}^{RF}}\left( {{{\bf{F}}^{\rm{H}}} \otimes {{\bf{I}}_{{N_{RF}}}}} \right){{\bf{W}}^{BB}}} \right) + 4{\rho _2}\sum\limits_{k = 1}^K {\sum\limits_{n = 1}^{{N_c}} {\nabla P_{k,n}^{\bf{W}}} }.
\end{array}
\end{equation}
where ${\nabla P_{k,n}^{\bf{W}}}$ is given by (\ref{EucGradP_W^k}), and ${{\bf{e}}_{k,n}} \in {\mathbb{R}^{K \times 1}}$ has all-zero entries except for its $k$-th entry, which is equal to 1.
\begin{figure*}[ht]
\begin{equation}\label{EucGradP_W^k}
{\nabla P_{k,n}^{\bf{W}}} = \left\{ {\begin{array}{*{20}{c}}
		{0,}&{\rm{if }}\left( {{{\gamma '}_{k,n}} - \sigma _c^2{\Gamma _{k,n}}} \right) \ge 0,\\
		{\left( {{{\gamma '}_{k,n}} - \sigma _c^2{\Gamma _{k,n}}} \right){{\bf{B}}_{k,n}}\left( {\left( {1 + {\Gamma _{k,n}}} \right){{\bf{w}}_{k,n}}{\bf{e}}_{k,n}^H - {\Gamma _{k,n}}{{\bf{W}}_n}} \right)}  ,&{{\rm{otherwise}}.}
\end{array}} \right.
\end{equation}
\end{figure*}

Under the RCG-based algorithm, the auxiliary variable updates according to the descent direction and the step size in each iteration. For the $q$-th iteration, the search direction ${\bf{\Pi }}_{\bf{W}}^{(q)}$ is determined by the Riemannian gradient  ${\rm{grad}}\left( {f_{\bf{W}}^{(q)}} \right)$ and the $(q-1)$-th search direction ${\bf{\Pi }}_{\bf{W}}^{(q-1)}$, namely
\begin{equation}\label{SearchD}
{\bf{\Pi }}_{\bf{W}}^{(q)} =  - {\rm{grad}}\left( {f_{\bf{W}}^{(q)}} \right) + \varpi _{\bf{W}}^{(q - 1)}{\rm{Proj}}_{\bf{W}}^{^{(q)}}\left( {{\bf{\Pi }}_{\bf{W}}^{(q - 1)}} \right).
\end{equation}
where $\varpi _{\bf{W}}^{(q)}$ is obtained by the Polak-Ribière formula~\cite{bManifold2}, expressed as
\begin{equation}\label{Polak-R}
\varpi _{\bf{W}}^{(q)} = \frac{{\left\langle {{\rm{grad}}\left( {f_{\bf{W}}^{(q)}} \right),{\rm{grad}}\left( {f_{\bf{W}}^{(q)}} \right) - {\rm{Proj}}_{\bf{W}}^{^{(q - 1)}}\left( {f_{\bf{W}}^{(q - 1)}} \right)} \right\rangle }}{{\left\langle {{\rm{grad}}\left( {f_{\bf{W}}^{(q - 1)}} \right),{\rm{grad}}\left( {f_{\bf{W}}^{(q - 1)}} \right)} \right\rangle }}.
\end{equation}
where $\left\langle { \cdot , \cdot } \right\rangle $ is the Euclidean inner product operation.

For the $q$-th iteration, the step size $\mu _{\bf{W}}^{(q)}$ is obtained by the Armijo line search rule~\cite{bManifold2}. With the descent direction  ${\bf{\Pi }}_{\bf{W}}^{(q)}$ and the step size $\mu _{\bf{W}}^{(q)}$ at the $q$-th iteration, the auxiliary variable is updated at the $(q+1)$-th iteration by the \textit{retraction} over the manifold $\mathcal{S_{\bf{W}}}$, namely
\begin{equation}\label{Ite-Auxil}
{{\bf{W}}^{(q + 1)}} = {\mathcal R}_{\bf{W}}^{(q)}\left( {\mu _{\bf{W}}^{(q)}{\bf{\Pi }}_{\bf{W}}^{(q)}} \right) = \frac{{\left( {{{\bf{W}}^{(q)}} + \mu _{\bf{W}}^{(q)}{\bf{\Pi }}_{\bf{W}}^{(q)}} \right)}}{{{{\left\| {{{\bf{W}}^{(q)}} + \mu _{\bf{W}}^{(q)}{\bf{\Pi }}_{\bf{W}}^{(q)}} \right\|}_{\rm{F}}}}}.
\end{equation}
where ${{\mathcal R}_{\bf{W}}}\left( {{\mu _{\bf{W}}}{{\bf{\Pi }}_{\bf{W}}}} \right) = {{\left( {{\bf{W}} + {\mu _{\bf{W}}}{{\bf{\Pi }}_{\bf{W}}}} \right)} \mathord{\left/{\vphantom {{\left( {{\bf{W}} + {\mu _{\bf{W}}}{{\bf{\Pi }}_{\bf{W}}}} \right)} {{{\left\| {{\bf{W}} + {\mu _{\bf{W}}}{{\bf{\Pi }}_{\bf{W}}}} \right\|}_F}}}} \right.	\kern-\nulldelimiterspace} {{{\left\| {{\bf{W}} + {\mu _{\bf{W}}}{{\bf{\Pi }}_{\bf{W}}}} \right\|}_{\rm{F}}}}}$ is the retraction operator, mapping a vector from the tangent space to the manifold $\mathcal{S_{\bf{W}}}$.

The whole procedure of the RCG-based algorithm for obtaining the auxiliary variable $\bf{W}$ is concluded as Algorithm \ref{AlgoRCG}. 
\begin{algorithm}[!t]
	\caption{The RCG-based algorithm for the auxiliary variable}
	\label{AlgoRCG}
	\begin{algorithmic}[1]
		\STATE \textbf{Input}:${{{\bf{W}}^{RF}}^{(t - 1)}},{{{\bf{W}}^{BB}}^{(t - 1)}},{{\Theta}^{(t - 1)}},{{\bf{\Lambda }}^{(t - 1)}},\rho_1,{Q_{\max }},$ and ${\varepsilon _{\bf{W}}} $. 
		\STATE \textbf{Initialize}: Set ${\bf{W}} \in \mathcal{S_{\bf{W}}}$ randomly, $q=1$.
		\REPEAT 
		\STATE Obtain the stepsize $\mu _{\bf{W}}^{(q)}$ by the Armijo rule.
		\STATE Obtain the Polak-Ribière parameter $\varpi _{\bf{W}}^{(q)}$ by (\ref{Polak-R}).
		\STATE Obtain the search direction ${\bf{\Pi }}_{\bf{W}}^{(q)}$ by (\ref{SearchD}).
		\STATE Update the auxiliary variable ${{\bf{W}}^{(q)}}$ by (\ref{Ite-Auxil}).
		\STATE Update $q=q+1$
		\UNTIL $q > {Q_{\max }}$ or ${\left\| {{\rm{grad}}\left( {{\rm{f}}_{\bf{W}}^{(q)}} \right)} \right\|_{\rm{F}}} \le {\varepsilon _{\bf{W}}}$\\		
		\STATE \textbf{Output}: the solved auxiliary variable ${{\bf{W}}^{(t)}} = {{\bf{W}}^{(q)}}$. \\
	\end{algorithmic}
\end{algorithm}

\subsection{Update of RF Beamforming Variable}
Given $\{ {\bf{W}},{{\bf{W}}^{BB}},{\Theta},{\bf{\Lambda }}\}$, the sub-problem with respect to the RF beamformer ${{\bf{W}}^{RF}}$ is expressed as
\begin{subequations}\label{OptRF}
	\begin{align}		
    {{{\bf{W}}^{RF}}^{(t)}} & =  \arg \mathop {\min }\limits_{{{\bf{W}}^{RF}}} L({{\bf{W}}^{(t)}},{{\bf{W}}^{RF}},{{{{\bf{W}}^{BB}}}^{(t - 1)}},{{\Theta}^{(t - 1)}},{{\bf{\Lambda }}^{(t - 1)}})\\
	\mbox{s.t.}\quad
	&\left| {{{{\bf{\tilde W}}}^{RF}}(i,j)} \right| = 1,\forall i \in {\mathcal{N}_t},j \in {\mathcal{N}_{RF}}.
	\end{align}
\end{subequations}
Omitting the unrelated terms, the RF analog beamforming problem becomes
\begin{subequations}\label{OptRF2}
	\begin{align}		
		\mathop {\min }\limits_{{{\bf{W}}^{RF}}} \frac{{{\rho _1}}}{2} & \left\| {{\bf{W}} + \frac{{\bf{\Lambda }}}{{{\rho _1}}} - {{\bf{W}}^{RF}}\left( {{{\bf{F}}^{\rm{H}}} \otimes {{\bf{I}}_{{N_{RF}}}}} \right){{\bf{W}}^{BB}}} \right\|_{\rm{F}}^2 \label{RFobj}\\
		\mbox{s.t.}\quad
		&\left| {{{{\bf{\tilde W}}}^{RF}}(i,j)} \right| = 1,\forall i \in {\mathcal{N}_t},j \in {\mathcal{N}_{RF}}.
	\end{align}
\end{subequations}
Since the RF analog beamformer is identical for each sub-carrier, we can reformulate the objective function in (\ref{OptRF2}) as follows
\begin{subequations}\label{OptRF3}
	\begin{align}		
		\mathop {\min }\limits_{{{{\bf{\tilde W}}}^{RF}}} \frac{{{\rho _1}}}{2}&\sum\limits_{n = 1}^{{N_c}} {\left\| {{{\bf{W}}_n} + \frac{{{{\bf{\Lambda }}_n}}}{{{\rho _1}}} - {{\bf{\tilde W}}^{RF}}{\left( {\left( {{{\bf{F}}^{\rm{H}}} \otimes {{\bf{I}}_{{N_{RF}}}}} \right){{\bf{W}}^{BB}}} \right)_n}} \right\|_{\rm{F}}^2} \label{WRFObj} \\
		\mbox{s.t.}\quad
		&\left| {{{{\bf{\tilde W}}}^{RF}}(i,j)} \right| = 1,\forall i \in {\mathcal{N}_t},j \in {\mathcal{N}_{RF}}.
	\end{align}
\end{subequations}

Similarly, the problem (\ref{OptRF3}) can also be regarded as a problem without constraint over the complex circle manifold ${{\mathcal S}_{{{{\bf{\tilde W}}}^{RF}}}} = \left\{ {{{{\bf{\tilde W}}}^{RF}} \in {\mathbb{C}^{{N_t} \times {N_{RF}}}}\left| {\left| {{{{\bf{\tilde W}}}^{RF}}(i,j)} \right| = 1} \right.} \right\}$, where the feasible set is a ${{{N}_t}\times K}-1$ dimension complex hypersphere. Therefore, the RF analog beamforming sub-problem can also be solved by the RCG-based algorithm, with a procedure similar to Algorithm \ref{AlgoRCG} \footnote{The procedure is omitted due to space constraint.}. The Euclidean gradient of (\ref{WRFObj}) towards ${{\bf{\tilde W}}^{RF}}$ is calculated as
\begin{equation}\label{EucGradWRF}
\begin{array}{l}
	\nabla {f_{{{{\bf{\tilde W}}}^{RF}}}} = {\rho _1}\sum\limits_{n = 1}^{{N_c}} {\left( {{{{\bf{\tilde W}}}^{RF}}{{\left( {\left( {{{\bf{F}}^{\rm{H}}} \otimes {{\bf{I}}_{{N_{RF}}}}} \right){{\bf{W}}^{BB}}} \right)}_n}{{\left( {{{\left( {\left( {{{\bf{F}}^{\rm{H}}} \otimes {{\bf{I}}_{{N_{RF}}}}} \right){{\bf{W}}^{BB}}} \right)}_n}} \right)}^{\rm{H}}}} \right.} \\
	\left. { - \left( {{{\bf{W}}_n} + \frac{{{{\bf{\Lambda }}_n}}}{{{\rho _1}}}} \right){{\left( {{{\left( {\left( {{{\bf{F}}^{\rm{H}}} \otimes {{\bf{I}}_{{N_{RF}}}}} \right){{\bf{W}}^{BB}}} \right)}_n}} \right)}^{\rm{H}}}} \right).
\end{array}
\end{equation}

\subsection{Update of Baseband Beamforming Variable}
Given $\{ {\bf{W}},{{\bf{W}}^{RF}},{\Theta},{\bf{\Lambda }}\}$, the baseband digital beamformer ${{\bf{W}}^{BB}}$ can be optimized by solving the following sub-problem:
\begin{equation}\label{OptBB}
{{{\bf{W}}^{BB}}^{(t)} = \arg \mathop {\min }\limits_{{{\bf{W}}^{BB}}} L({{\bf{W}}^{(t)}},{{\bf{W}}^{RF}}^{(t)},{{\bf{W}}^{BB}},{{\Theta}^{(t - 1)}},{{\bf{\Lambda }}^{(t - 1)}})}.
\end{equation}
Omitting the unrelated parameters, the baseband digital beamforming problem becomes
\begin{equation}\label{OptBB2}
\mathop {\min }\limits_{{{\bf{W}}^{BB}}} \frac{{{\rho _1}}}{2}\left\| {{\bf{W}} + \frac{{\bf{\Lambda }}}{{{\rho _1}}} - {{\bf{W}}^{RF}}\left( {{{\bf{F}}^{\rm{H}}} \otimes {{\bf{I}}_{{N_{RF}}}}} \right){{\bf{W}}^{BB}}} \right\|_{\rm{F}}^2.
\end{equation}
With no constraints in (\ref{OptBB2}), the closed form of ${{\bf{W}}^{BB}}$ can be obtained as follows
\begin{equation}\label{WBB}
{\left( {{{\bf{W}}^{BB}}} \right)^*} = {\left( {{{\left( {{{\bf{W}}^{RF}}\left( {{{\bf{F}}^{\rm{H}}} \otimes {{\bf{I}}_{{N_{RF}}}}} \right)} \right)}^{\rm{H}}}{{\bf{W}}^{RF}}\left( {{{\bf{F}}^{\rm{H}}} \otimes {{\bf{I}}_{{N_{RF}}}}} \right)} \right)^{ - 1}}{\left( {{{\bf{W}}^{RF}}\left( {{{\bf{F}}^{\rm{H}}} \otimes {{\bf{I}}_{{N_{RF}}}}} \right)} \right)^{\rm{H}}}\left( {{\bf{W}} + \frac{{\bf{\Lambda }}}{{{\rho _1}}}} \right).
\end{equation}

\subsection{Update of Phase Shift Variable}
Given $\{ {\bf{W}},{{\bf{W}}^{RF}},{{\bf{W}}^{BB}},{\bf{\Lambda }}\}$, the problem for the phase shift variable ${\Theta}$ is expressed as
\begin{equation}\label{OptPSV}
{{\Theta}^{(t)} = \arg \mathop {\min }\limits_{\Theta} L({{\bf{W}}^{(t)}},{{\bf{W}}^{RF}}^{(t)},{{\bf{W}}^{BB}}^{(t)},{\Theta},{{\bf{\Lambda }}^{(t - 1)}})}.
\end{equation}
To decouple the phase shift variable with all the other variables and parameters, the transformation is first performed as $({\bf{h}}_{bu,kn}^{\rm{H}} + {\bf{h}}_{ru,kn}^{\rm{H}}\Theta {\bf{H}}_{br,n}^{\rm{H}}){{\bf{w}}_{k,n}} = {b_{k,n}} + {{\bf{v}}^{\rm{H}}}{{\bf{a}}_{k,n}}$ in the SINR constraint, where ${\bf{v}} = {[{e^{j{\theta _1}}},...,{e^{j{\theta _r}}},...,{e^{j{\theta _R}}}]^{\rm{H}}}$. ${v_r} = {e^{j{\theta _r}}},\forall r$. ${{\bf{a}}_{k,n}} = {\rm{diag}}({\bf{h}}_{ru,kn}^{\rm{H}}){\bf{H}}_{br,n}^{\rm{H}}{{\bf{w}}_{k,n}}$ and ${b_{k,n}} = {\bf{h}}_{bu,kn}^{\rm{H}}{{\bf{w}}_{k,n}}$. For the objective function, only the first part of (\ref{ALag}) is related to ${\Theta}$. Therefore, we omit the unrelated second part and transform the first part of the objective function for each sub-carrier as follows
\begin{equation}\label{OptPSVtheta}
\begin{array}{l}
	\sum\limits_\psi ^{{N_\psi }} {{{\left| {{{\bf{A}}^{\rm{H}}}\left( \psi  \right)\hat \Theta {\bf{H}}_{br}^{\rm{H}}{\bf{W}}{{\bf{W}}^{\rm{H}}}{{\bf{H}}_{br}}{{\hat \Theta }^{\rm{H}}}{\bf{A}}\left( \psi  \right) - {\rm{B}}{{\rm{P}}_{{\rm{ref}}}}\left( \psi  \right)} \right|}^2}} \\
	= \sum\limits_\psi ^{{N_\psi }} {\sum\limits_n^{{N_c}} {{{\left| {{{\bf{v}}^{\rm{H}}}{\rm{diag}}\left( {{\bf{\hat a}}_n^{\rm{H}}\left( \psi  \right)} \right){\bf{H}}_{br,n}^{\rm{H}}{{\bf{W}}_n}{\bf{W}}_n^{\rm{H}}{{\bf{H}}_{br,n}}{\rm{diag}}\left( {{{{\bf{\hat a}}}_n}\left( \psi  \right)} \right){\bf{v}} - {\rm{B}}{{\rm{P}}_{{\rm{ref}},{\rm{n}}}}\left( \psi  \right)} \right|}^2}} }.
\end{array}
\end{equation}
Denoting ${\rm{diag}}\left( {{\bf{\hat a}}_n^{\rm{H}}\left( \psi  \right)} \right){\bf{H}}_{br,n}^{\rm{H}}{{\bf{W}}_n}$ as ${{\bf{C}}_{\psi n}}$, the decoupled sub-problem for the phase shift variable ${\bf{v}}$ turns into the following problem ${\rm P}3$, given by
\begin{subequations}\label{P5OptPS}
	\begin{align}
		({\rm P}3) \mathop {\min }\limits_{\bf{v}} & \sum\limits_\psi ^{{N_\psi }} {\sum\limits_n^{{N_c}} {{{\left| {{{\bf{v}}^{\rm{H}}}{{\bf{C}}_{\psi n}}{\bf{C}}_{\psi n}^{\rm{H}}{\bf{v}} - {\rm{B}}{{\rm{P}}_{{\rm{ref,n}}}}\left( \psi  \right)} \right|}^2}} } \\
		\mbox{s.t.}\quad
		&\frac{{{{\left| {{b_{k,n}} + {{\bf{v}}^{\rm{H}}}{{\bf{a}}_{k,n}}} \right|}^2}}}{{\sum\limits_{i = 1,i \ne k}^K {{{\left| {{b_{i,n}} + {{\bf{v}}^{\rm{H}}}{{\bf{a}}_{i,n}}} \right|}^2}}  + \sigma _c^2}} \ge {\Gamma _{k,n}},\forall k \in \mathcal{K},n \in \mathcal{N}_c, \label{SINRv3}\\
		&\left| {{v_r}} \right| = 1,\forall r\in {\cal R} \label{ConsPSv5},
	\end{align}
\end{subequations}
where ${b_{i,n}} = {\bf{h}}_{bu,kn}^{\rm{H}}{{\bf{w}}_{i,n}}$ and ${{\bf{a}}_{i,n}} = {\rm{diag}}({\bf{h}}_{ru,kn}^{\rm{H}}){\bf{H}}_{br,n}^{\rm{H}}{{\bf{w}}_{i,n}}$. The SINR constraint (\ref{SINRv3}) is still a challenging part for solving ${\bf{v}}$. To ensure a near-optimal and fast solution for the problem ${\rm P}3$, we first transform the SINR constraint and then utilize the similar RCG-based algorithm as Algorithm \ref{AlgoRCG}. The SINR constraint (\ref{SINRv3}) can be equivalently transformed as follows
\begin{equation}\label{SINRvTrans}
	\left( {\left( {1 + {\Gamma _{k,n}}} \right){{\left| {{b_{k,n}} + {{\bf{v}}^{\rm{H}}}{{\bf{a}}_{k,n}}} \right|}^2} - {\Gamma _{k,n}}\sum\limits_{i = 1}^K {{{\left| {{b_{i,n}} + {{\bf{v}}^{\rm{H}}}{{\bf{a}}_{i,n}}} \right|}^2}}  - \sigma _c^2{\Gamma _{k,n}}} \right) \ge 0,\forall k \in {\cal K},n \in {{\cal N}_c}.
\end{equation}
Leveraging the penalty method, the SINR constraint (\ref{SINRvTrans}) is integrated into the objective function turning into the problem ${\rm P}3.1$ as follows
\begin{subequations}\label{OptPStransform5}
	\begin{align}
		({\rm P}3.1) \quad \mathop {\min }\limits_{\bf{v}} {f_{\bf{v}}} & = \sum\limits_\psi ^{{N_\psi }} {\sum\limits_n^{{N_c}} {{{\left| {{{\bf{v}}^{\rm{H}}}{{\bf{C}}_{\psi n}}{\bf{C}}_{\psi n}^{\rm{H}}{\bf{v}} - {\rm{B}}{{\rm{P}}_{{\rm{ref,n}}}}\left( \psi  \right)} \right|}^2}} }  + {\rho _3}\sum\limits_{k = 1}^K {\sum\limits_{n = 1}^{{N_c}} {{P_{k,n}}({\bf{v}})} } \label{Objv31} \\
		\mbox{s.t.}\quad
		&\left| {{v_r}} \right| = 1,\forall r\in {\cal R}.
	\end{align}
\end{subequations}
where ${{P_{k,n}}({\bf{v}})}$ is given by
\begin{equation}\label{PenaltyV}
	{P_{k,n}}({\bf{v}}) = {\left( {\min \left\{ {\left( {\left( {1 + {\Gamma _{k,n}}} \right){{\left| {{b_{k,n}} + {{\bf{v}}^{\rm{H}}}{{\bf{a}}_{k,n}}} \right|}^2} - {\Gamma _{k,n}}\sum\limits_{i = 1}^K {{{\left| {{b_{i,n}} + {{\bf{v}}^{\rm{H}}}{{\bf{a}}_{i,n}}} \right|}^2}}  - \sigma _c^2{\Gamma _{k,n}}} \right),0} \right\}} \right)^2}.
\end{equation}
Then, the problem ${\rm P}3.1$ can be regarded as a problem without constraint over the complex circle manifold $\mathcal{S}_{\bf{v}} = \left\{ {{\bf{v}} \in {\mathbb{C}^{R \times 1}}\left| {\left| {{v_r}} \right| = 1,\forall r \in \mathcal{R}} \right.} \right\}$, which can be solved by the RCG-based algorithm, similar to the Algorithm \ref{AlgoRCG}, thus not repeated here. The Euclidean gradient of (\ref{Objv31}) is given by
\begin{equation}\label{EucGradv31}
	\nabla {f_{\bf{v}}} = 4\sum\limits_\psi ^{{N_\psi }} {\sum\limits_n^{{N_c}} {\left( {{{\bf{C}}_{\psi n}}{\bf{C}}_{\psi n}^{\rm{H}}{\bf{v}}{{\bf{v}}^{\rm{H}}}{{\bf{C}}_{\psi n}}{\bf{C}}_{\psi n}^{\rm{H}}{\bf{v}} - {{\bf{C}}_{\psi n}}{\bf{C}}_{\psi n}^{\rm{H}}{\bf{v}}{\rm{B}}{{\rm{P}}_{{\rm{ref,n}}}}\left( \psi  \right)} \right)} }  + {\rho _3}\sum\limits_{k = 1}^K {\sum\limits_{n = 1}^{{N_c}} {\nabla {P_{k,n}}({\bf{v}})} }.
\end{equation}
where ${\nabla {P_{k,n}}({\bf{v}})}$ is derived as (\ref{EucGradv2}). 

\begin{figure*}[ht]
\begin{footnotesize}
	\begin{equation}\label{EucGradv2}
		\begin{array}{l}
			{\nabla {P_{k,n}}({\bf{v}})} = \\
			\left\{ {\begin{array}{*{20}{c}}
					{0,}&{\begin{array}{*{20}{l}}
							{{\rm{if }}\left( {\left( {1 + {\Gamma _{k,n}}} \right){{\left| {{b_{k,n}} + {{\bf{v}}^{\rm{H}}}{{\bf{a}}_{k,n}}} \right|}^2}} \right.}\\
							{\left. { - {\Gamma _{k,n}}\sum\limits_{i = 1}^K {{{\left| {{b_{i,n}} + {{\bf{v}}^{\rm{H}}}{{\bf{a}}_{i,n}}} \right|}^2}}  - \sigma _c^2{\Gamma _{k,n}}} \right) \ge 0,}
					\end{array}}\\
					{\begin{array}{*{20}{c}}
							{4\left( {\left( {1 + {\Gamma _{k,n}}} \right){{\left| {{b_{k,n}} + {{\bf{v}}^{\rm{H}}}{{\bf{a}}_{k,n}}} \right|}^2}\left. { - {\Gamma _{n,k}}\sum\limits_{i = 1}^K {{{\left| {{b_{i,n}} + {{\bf{v}}^{\rm{H}}}{{\bf{a}}_{i,n}}} \right|}^2}}  - \sigma _c^2{\Gamma _{k,n}}} \right) \times } \right.}\\
							{\left( {\left( {1 + {\Gamma _{k,n}}} \right)\left( {{{\bf{a}}_{k,n}}{\bf{a}}_{k,n}^{\rm{H}}{\bf{v}} + {{\bf{a}}_{k,n}}b_{k,n}^{\rm{H}}} \right) - {\Gamma _{k,n}}\sum\limits_{i = 1}^K {\left( {{{\bf{a}}_{i,n}}{\bf{a}}_{i,n}^{\rm{H}}{\bf{v}} + {{\bf{a}}_{i,n}}b_{i,n}^{\rm{H}}} \right)} } \right)}
					\end{array}}&{{\rm{otherwise}}.}
			\end{array}} \right.
		\end{array}
	\end{equation}
	\end{footnotesize}
\end{figure*}

\subsection{Update of Lagrangian Multiplier}
The update rule of Lagrangian multipliers ${\bf{\Lambda }}$ based on the idea of dual ascent, is expressed as
\begin{equation}\label{OptLM}
{{\bf{\Lambda }}^{(t)}} = {{\bf{\Lambda }}^{(t - 1)}} + {\rho _1}\left( {{{\bf{W}}^{(t)}} - {{\bf{W}}^{RF}}^{(t)}\left( {{{\bf{F}}^{\rm{H}}} \otimes {{\bf{I}}_{{N_{RF}}}}} \right){{\bf{W}}^{BB}}^{(t)}} \right).
\end{equation}

In conclusion, all the variables and Lagrangian multipliers $\{ {\bf{W}},{{\bf{W}}^{RF}},{{\bf{W}}^{BB}},{\Theta},{\bf{\Lambda }}\}$ are alternatingly iterated under the ADMM framework until the convergence or the maximal number of ADMM iteration ${I_{ADMM}}$ is reached. This manifold-based ADMM algorithm for solving problem ${\rm P}1$ is summarized as Algorithm \ref{AlgoBCD-ADMM}.
\begin{algorithm}[!t]
	\caption{Manifold-based ADMM algorithm for solving problem P1}
	\label{AlgoBCD-ADMM}
	\begin{algorithmic}[1]
		\STATE \textbf{Input}:$ {\bf{W}}$, ${{\bf{W}}^{RF}}$, ${{\bf{W}}^{BB}}$, ${\Theta}$, ${\bf{\Lambda }}$, $\rho_1$, $\rho_2$, $\rho_3$, and ${I_{ADMM}}$.\\
        \STATE \textbf{Initialize}: Set ${\bf{W}},{{\bf{W}}^{RF}},{{\bf{W}}^{BB}}$ randomly, ${\rho _1}=0.1$, ${\rho _2}=0.1$, ${\rho _3}=0.1$ and ${\bf{\Lambda }}=\textbf{0}$.
        \REPEAT
            \STATE Update the auxiliary variable ${\bf{W}}$ by Algorithm 1. \\
            \STATE Update the RF beamforming variable ${{\bf{W}}^{RF}}$ by RCG-based algorithm. \\
            \STATE Update the baseband beamforming variable ${{\bf{W}}^{BB}}$ by (\ref{WBB}). \\
            \STATE Update the phase shift variable ${\bf{v}}$ by RCG-based algorithm. \\
            \STATE Update the Lagrangian multipliers ${\bf{\Lambda }}$ by (\ref{OptLM}). \\
            \STATE Update the penalty parameters as $\min \left\{ {10{\rho _x},1000} \right\}$, $x = 1,2,3$. \\               
        \UNTIL convergence or ${I_{ADMM}}$ is reached. \\
        \STATE \textbf{Output}: the solved variables $\{ {\bf{W}},{{\bf{W}}^{RF}},{{\bf{W}}^{BB}},{\Theta}\}$. \\
	\end{algorithmic}
\end{algorithm}

\subsection{Complexity Analysis}
In this section, we analyze the computational complexity of the proposed algorithm, where the complexity mainly depends on obtaining ${\bf{W}}$,${{\bf{W}}^{RF}}$, ${{\bf{W}}^{BB}}$ and ${\Theta}$. Among these variables, the complexity of the manifold-based solutions for ${\bf{W}}$,${{\bf{W}}^{RF}}$ and ${\Theta}$ dominantly depends on calculating the Euclidean gradient, which are ${\mathcal O}( {{I_{\bf{W}}}\left( {{N_c}{N_\psi }N_t^2K + N_c^3N_{RF}^2{N_t}} \right)} )$, ${\mathcal O}({I_{{{\bf{W}}^{RF}}}}(N_c^4N_{RF}^2K + {N_c}N_t^2{K^2}{N_{RF}}))$ and ${\mathcal O}({I_\Theta }{N_c}{N_\psi }{R^2}K)$, respectively. ${I_{\bf{W}}}$, ${I_{{{\bf{W}}^{RF}}}}$ and ${I_\Theta }$ are the number of iterations for solving ${\bf{W}}$,${{\bf{W}}^{RF}}$ and ${\Theta}$, respectively. The complexity of calculating the closed-form solution for ${{\bf{W}}^{BB}}$ is ${\mathcal O}(N_c^3N_{RF}^2{N_t})$. Therefore, the total computational complexity of the proposed algorithm is ${\mathcal O}({I_{ADMM}}({I_{\bf{W}}}({N_c}{N_\psi }N_t^2K + N_c^3N_{RF}^2{N_t}) + {I_{{{\bf{W}}^{RF}}}}(N_c^4N_{RF}^2K + {N_c}N_t^2{K^2}{N_{RF}}) + {I_\Theta }{N_c}{N_\psi }{R^2}K + N_c^3N_{RF}^2{N_t}))$, where ${I_{ADMM}}$ is the number of iterations for ADMM algorithm.

\section{Numerical Results}
In this section, we evaluate our proposed manifold-based ADMM scheme via numerical simulation from four aspects. Firstly, the convergence of the proposed scheme is confirmed by observing that the objective function of problem ${\rm P}1$ stabilizes after some iterations. Then, to validate the sensing performance, the designed beampattern at the RIS obtained from the proposed algorithm is compared against the reference beampattern, and the beampatterns from the other three schemes. Furthermore, to measure the impact of different parameters on the system performance, we analyze the SINR feasibility ratio\footnote{The SINR feasibility ratio is defined as the percentage of satisfying SINR requirements by all UEs over all sub-carriers.} for communication, and the beampattern MSE and PSLR for sensing, considering different values of the number of UEs, RIS size, SINR threshold, and transmit SNR\footnote{The transmit SNR is defined as the ratio of transmit power to noise power at the transmitter side. We use SNR to represent it shortly in the rest of the paper.}. Finally, the performance tradeoff between sensing and communication is analyzed.

To assess the performance of our proposed scheme, we compare our solution against three benchmarks, namely,
\begin{itemize}
	\item \textbf{Manifold-based fully-digital beamforming with RIS (denoted as ‘FDB with RIS’)}: To obtain this scheme, we replace the hybrid beamforming with the fully-digital beamforming, resulting in an upper bound performance for the proposed solution. Here, the variables of digital beamformer and phase shifts are alternatingly optimized via the same penalty method and manifold optimization used for the proposed scheme.
	
	\item \textbf{Manifold-based fully-digital beamforming with random RIS (denoted as ‘FDB with rnd RIS’)}: The difference between this scheme and the ‘FDB with RIS’ scheme is that RIS is configured with random phase shifts.
	
	\item \textbf{Manifold-based hybrid beamforming with random RIS (denoted as ‘HB with rnd RIS’)}. In this scheme, the RIS is configured with random phase shifts while the optimization of hybrid beamforming is the same as the proposed scheme.
\end{itemize}

\begin{table}[!t]
	\caption{Simulation Parameters.}
	\label{Tab1}
	\setlength{\tabcolsep}{1pt} 
	\centering
	\begin{tabular}{c|c|c|c|c|c}
		\toprule 
		\textbf{Notation} & \textbf{Explanation} & \textbf{Value} & \textbf{Notation} & \textbf{Explanation} & \textbf{Value}  \\
		\midrule 
		$N_t$ & number of antennas & 64 & ${P_{\max }}$ & maximal power of BS & 30 dBm \\
		$N_{RF}$ & number of RF chains & 6-16 & SNR & ratio of transmit power over noise power & 19-31 dB \\
		$N_c$ & number of sub-carriers & 8 &  ${\Gamma _{n,k}}$ & SINR threshold of user $k$ on sub-carrier $n$ & 6-14 dB \\
		$f_c$ & center frequency & 28 GHz & ${Q_{\max }}$  & maximal iterations for manifold solution & 1000 \\
		$K$ & number of users & 2-10 &${\varepsilon _{\bf{W}}}$ & accuracy for convergence & 0.001 \\
		$M$ & number of targets & 3 &  ${\rho _1}$ & penalty parameter of ADMM problem & $\min \left\{ {10{\rho _1},1000} \right\}$ \\
		- & location of targets & [-50$^\circ$,0$^\circ$,50$^\circ$] &${\rho _2}$ & penalty parameter of SINR constraint for ${\bf{W}}$ & $\min \left\{ {10{\rho _2},1000} \right\}$ \\
		$R$ & number of RIS elements & 30-70 & ${\rho _3}$ & penalty parameter of SINR constraint for ${\bf{v}}$ & $\min \left\{ {10{\rho _3},1000} \right\}$ \\
		\bottomrule 
	\end{tabular}
\end{table}

For the \textbf{simulation setup} of the considered RIS-assisted mmWave OFDM JCAS system, we assume that the BS serves multiple UEs varying from $K=2$ to $K=10$ and senses three targets ($M=3$) located in the directions [$-50^\circ$,$0^\circ$,$50^\circ$]. The BS is configured with ${N_t}=64$ antennas and its transmit power is ${P_{\max }}=30$ dBm. The RIS is equipped with $R=30,40,50,60,70$ elements in different scenarios. The system works at center frequency ${f_c}=28$ GHz with ${N_c}=8$ sub-carriers. The number of RF chains $N_{RF}$ varies from $6$ to $16$. The SINR threshold ${\Gamma _{n,k}}$ increases from $6$ dB to $14$ dB. The SNR changes from $19$ dB to $31$ dB. The aforementioned parameters and other important parameters are summarized in Table ~\ref{Tab1}. Regarding the \textbf{initialization}, we initialize all the variables $ {\bf{W}},{{\bf{W}}^{RF}},{{\bf{W}}^{BB}}$ and ${\Theta}$ with random values. The penalty parameters ${\rho _1}$, ${\rho _2}$ and ${\rho _3}$ are all initialized as $0.1$ and increase every iteration by ten times till 1000. The Lagrangian multiplier matrix ${\bf{\Lambda }}$ is initialized as a zero matrix \textbf{0}.

\subsection{Convergence Performance}
\begin{figure}
	\centering
	\subfigure[Convergence under different channel realizations]{
		\begin{minipage}[b]{0.475\textwidth}
			\includegraphics[width=1\textwidth]{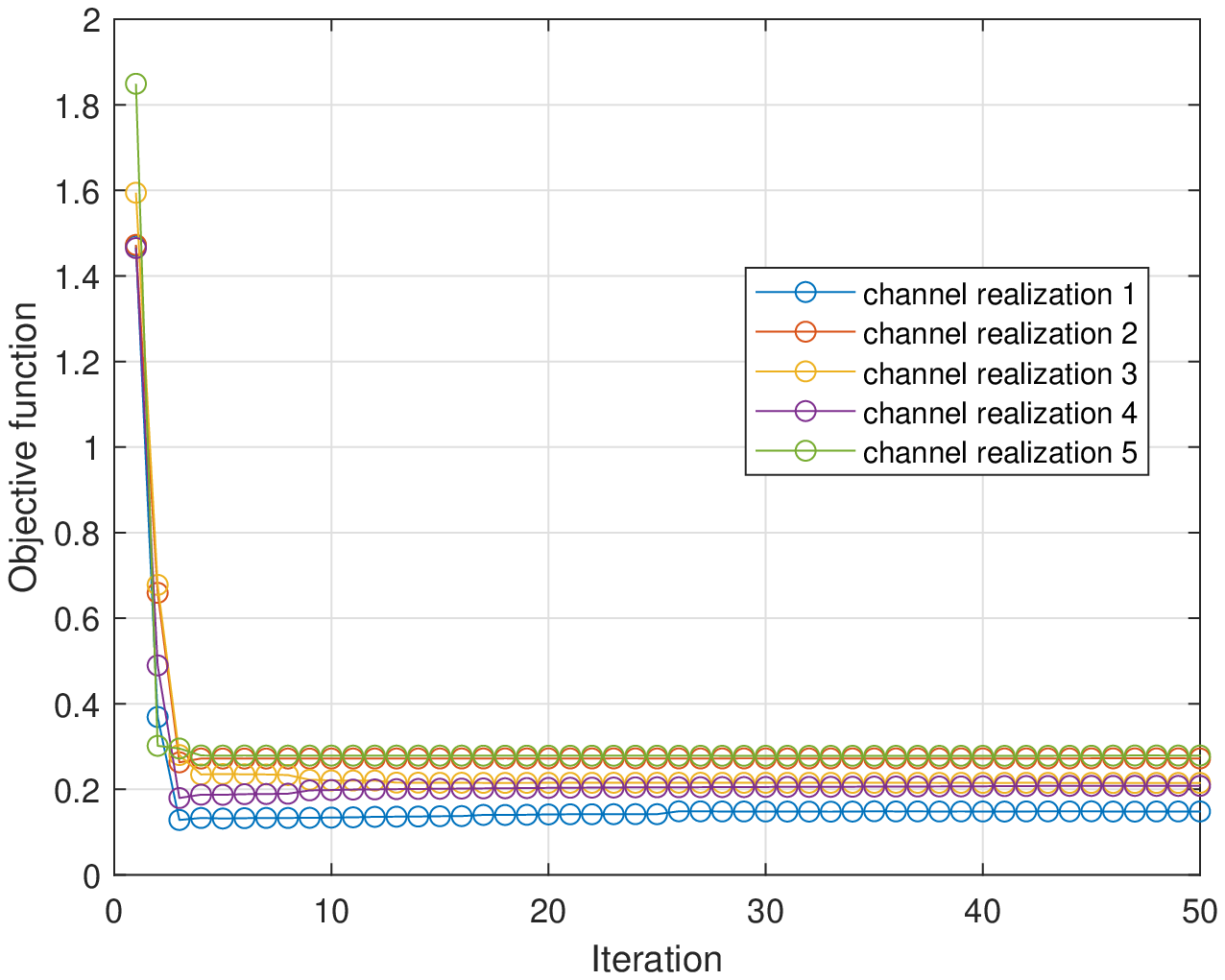}
		\end{minipage}
		\label{Conver_ObjCR}
	}
    	\subfigure[Convergence under varying number of RF chains]{
    		\begin{minipage}[b]{0.475\textwidth}
   		 	\includegraphics[width=1\textwidth]{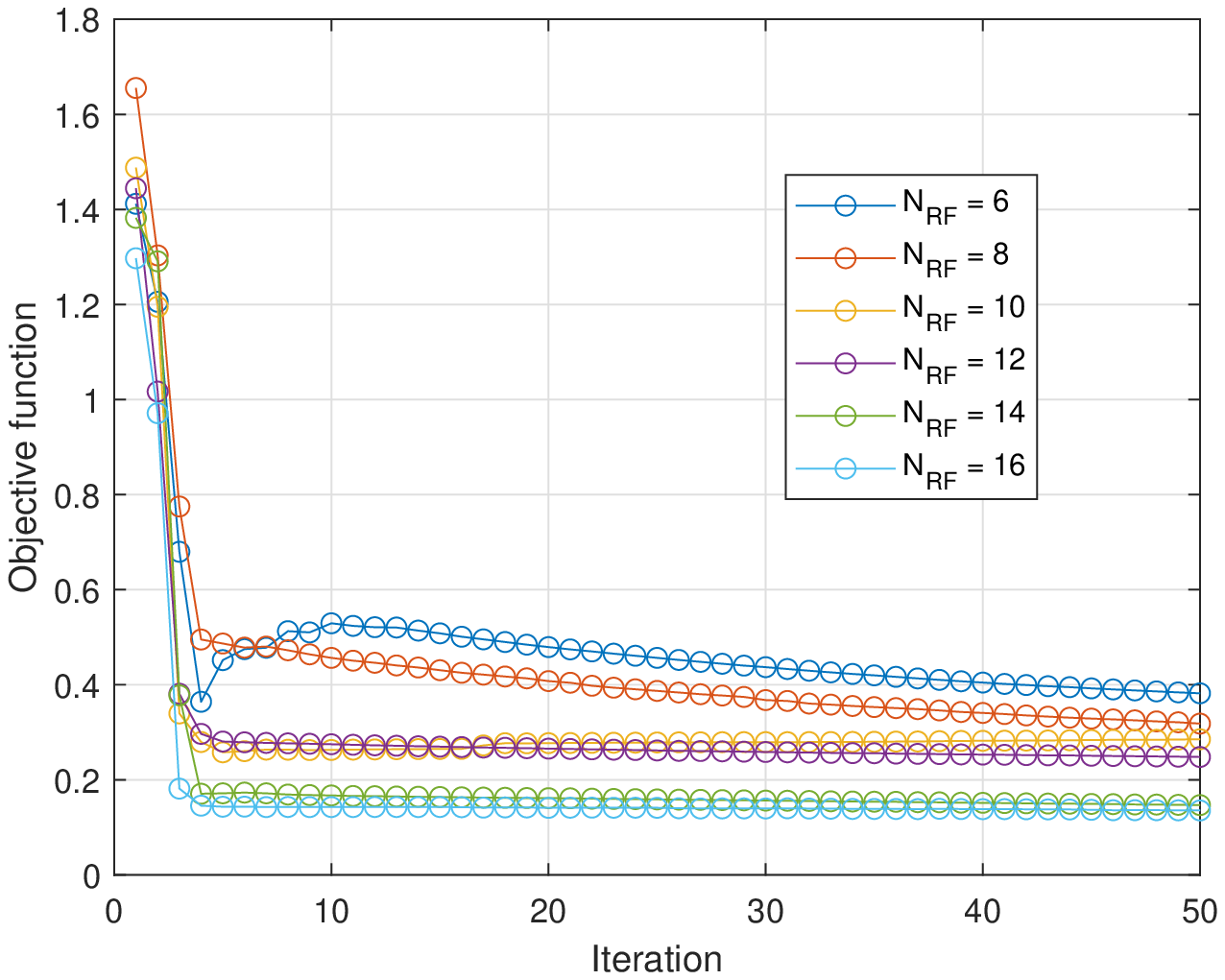}
    		\end{minipage}
		\label{Conver_ObjNRF}
    	}
	\caption{Convergence of the manifold-based ADMM algorithm}
	\label{Conver_Obj}
\end{figure}

In this sub-section, we evaluate the convergence of the proposed algorithm. In Fig.~\ref{Conver_Obj}, we observe that the objective function, namely the MSE between the designed and reference beampattern at the RIS, converges fast and approximates to zero under different channel realizations. Moreover, when the number of RF chains $N_{RF}$ increases, the objective function converges to a smaller value, indicating that the beampattern MSE can be decreased by adjusting parameters, e.g., $N_{RF}$. Therefore, Fig.~\ref{Conver_Obj} demonstrates that the proposed algorithm converges.

\subsection{Beampattern Performance}
\begin{figure}[!t]
\centering
  \begin{minipage}{0.5\linewidth}
    \centering
    \includegraphics[scale=0.55]{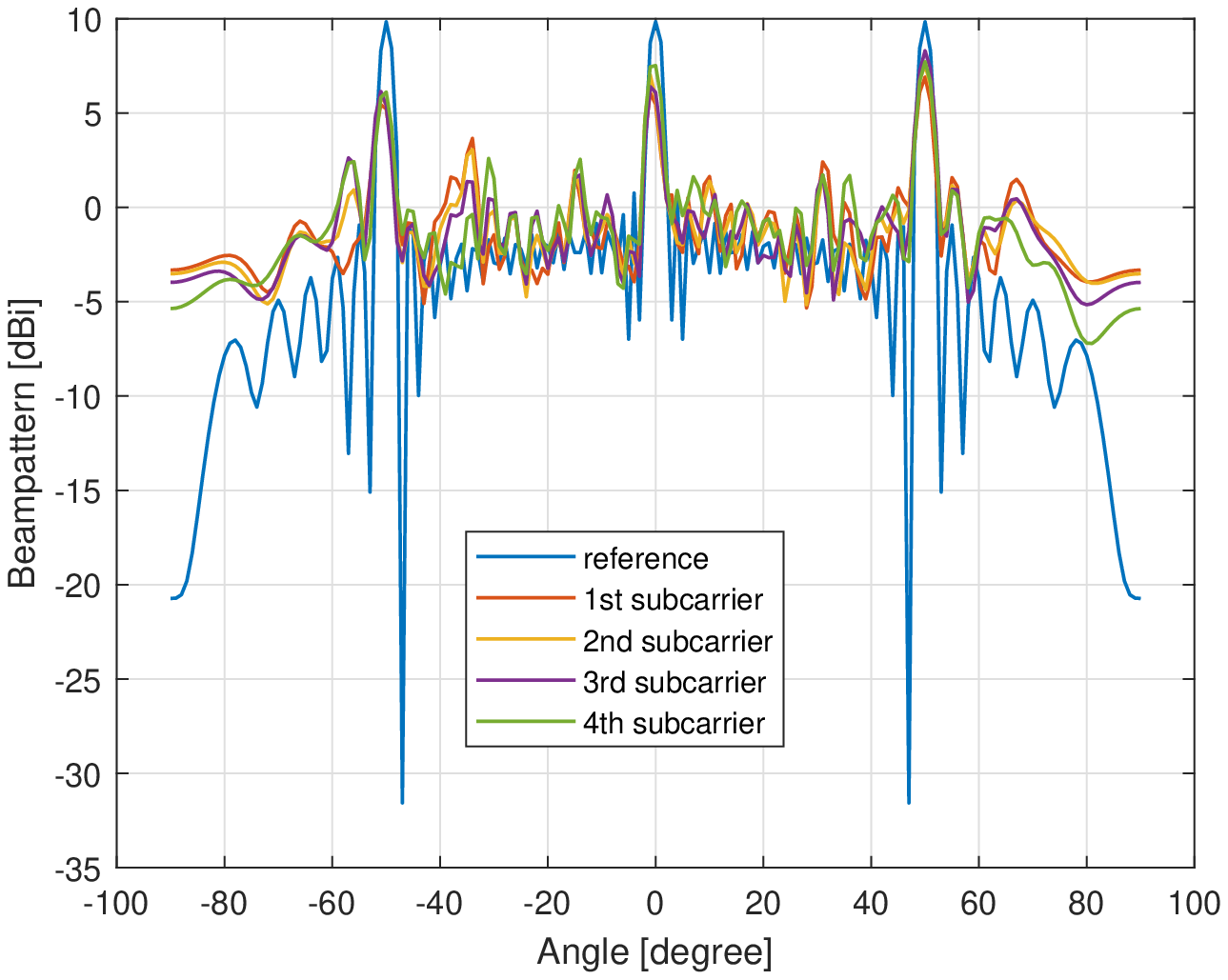}
  \end{minipage}%
  \begin{minipage}{0.5\linewidth}
    \centering
    \includegraphics[scale=0.55]{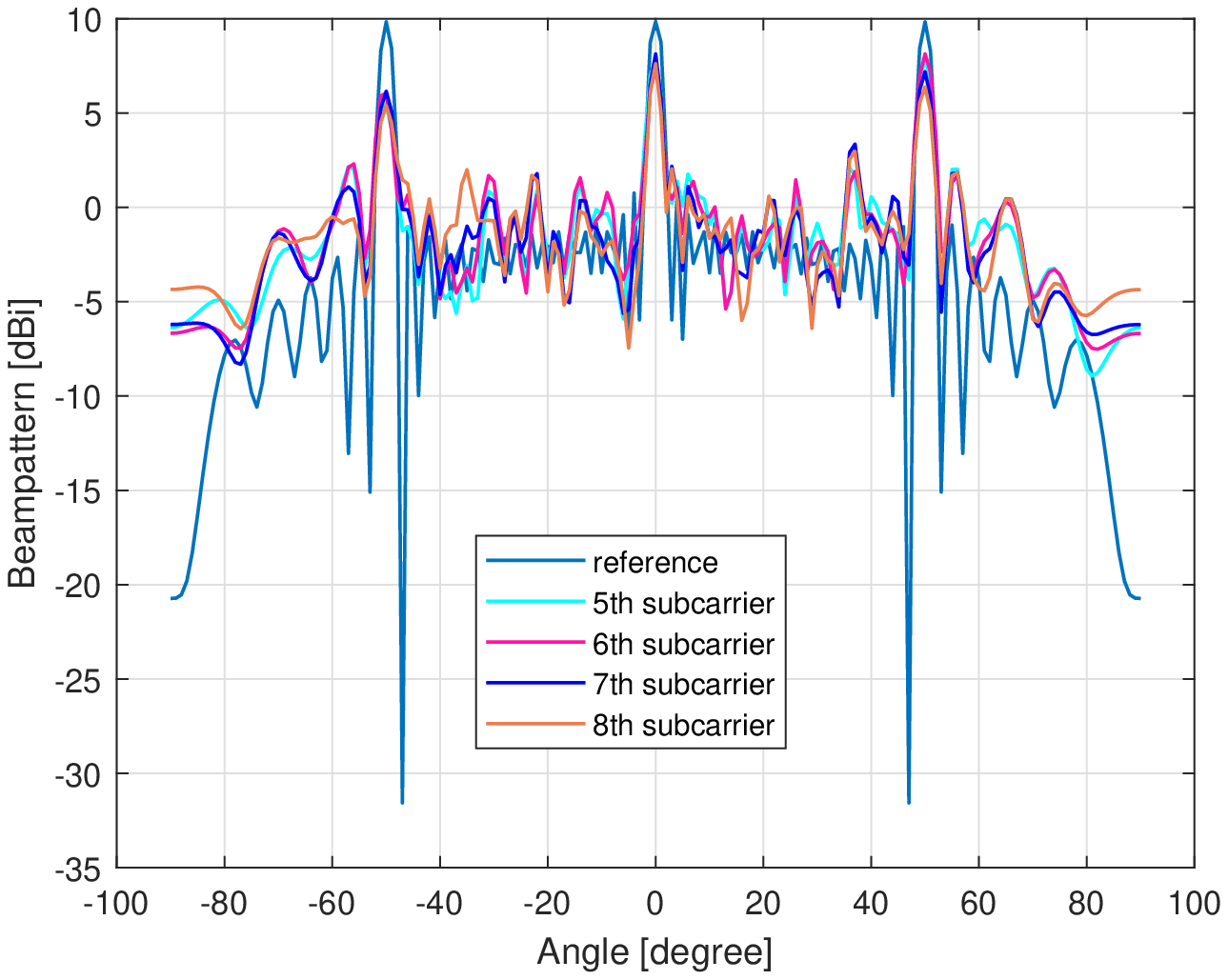}
  \end{minipage}
  \caption{Beampattern for each of the sub-carriers}
  \label{Beampattern8SC}
\end{figure}

In this sub-section, we illustrate the beampattern obtained by the proposed algorithm and compare it with the reference beampattern and the other three benchmark schemes to measure the sensing performance. Fig.~\ref{Beampattern8SC} shows the designed beampatterns across all the eight sub-carriers. The curve denoted as ‘reference’ represents the reference beampattern, which is identical for all sub-carriers. We observe that the designed beampatterns obtained by our proposed algorithm have main lobes pointing to the directions of interest. The gap of beampattern gain between the reference and designed beampattern exists because the reference beampattern is generated with the total transmit power only for sensing, while in the designed JCAS system, sensing and communication are both realized and only part of the transmit power is used for sensing.

\begin{figure}[!t]
	\centering
	\includegraphics[width=0.5\linewidth]{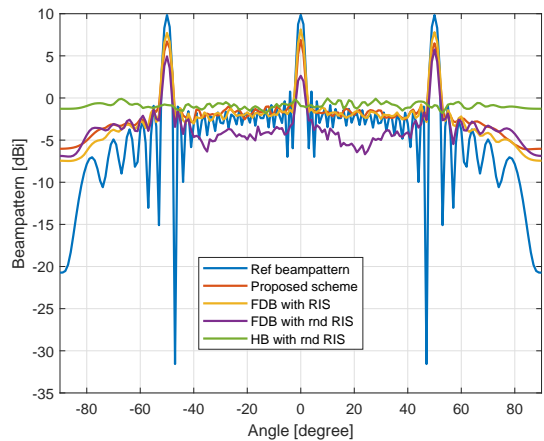}
	\caption{Beampatterns under different schemes}
	\label{BPcomK4}
\end{figure}

In Fig.~\ref{BPcomK4}, the beampatterns obtained from different schemes are compared, where the beampattern results are added up over all the sub-carriers. It can be observed that the fully-digital beamforming scheme (‘FDB with RIS’) has higher main lobes than the corresponding hybrid beamforming scheme (proposed scheme), showing a -10.8 dB beampattern MSE gap. This higher performance is reached at the cost of consuming ${N_{RF}}=N_t=64$ RF chains, which is much more than the number of RF chains ${N_{RF}}=16$ used in hybrid beamforming. Besides, the beampatterns obtained by the schemes with random RIS are worse than the corresponding schemes with optimized RIS. Particularly, the beampattern MSE between ‘FDB with RIS’ scheme and ‘FDB with rnd RIS’ scheme is -3.38 dB. The ‘HB with rnd RIS’ scheme does not have an obvious beampattern due to random RIS configuration.

\subsection{System Performance under Varying Parameters}
In this section, we evaluate the impact of the number of UEs, RIS size, SINR threshold, and SNR on the system performance. In particular, we use the SINR feasibility ratio for communication, and beampattern MSE or PSLR for sensing as a means of comparison.

\subsubsection{Impact of number of UEs}

\begin{figure}
	\centering
	\subfigure[SINR feasibility ratio]{
		\begin{minipage}[b]{0.475\textwidth}
			\includegraphics[width=1\textwidth]{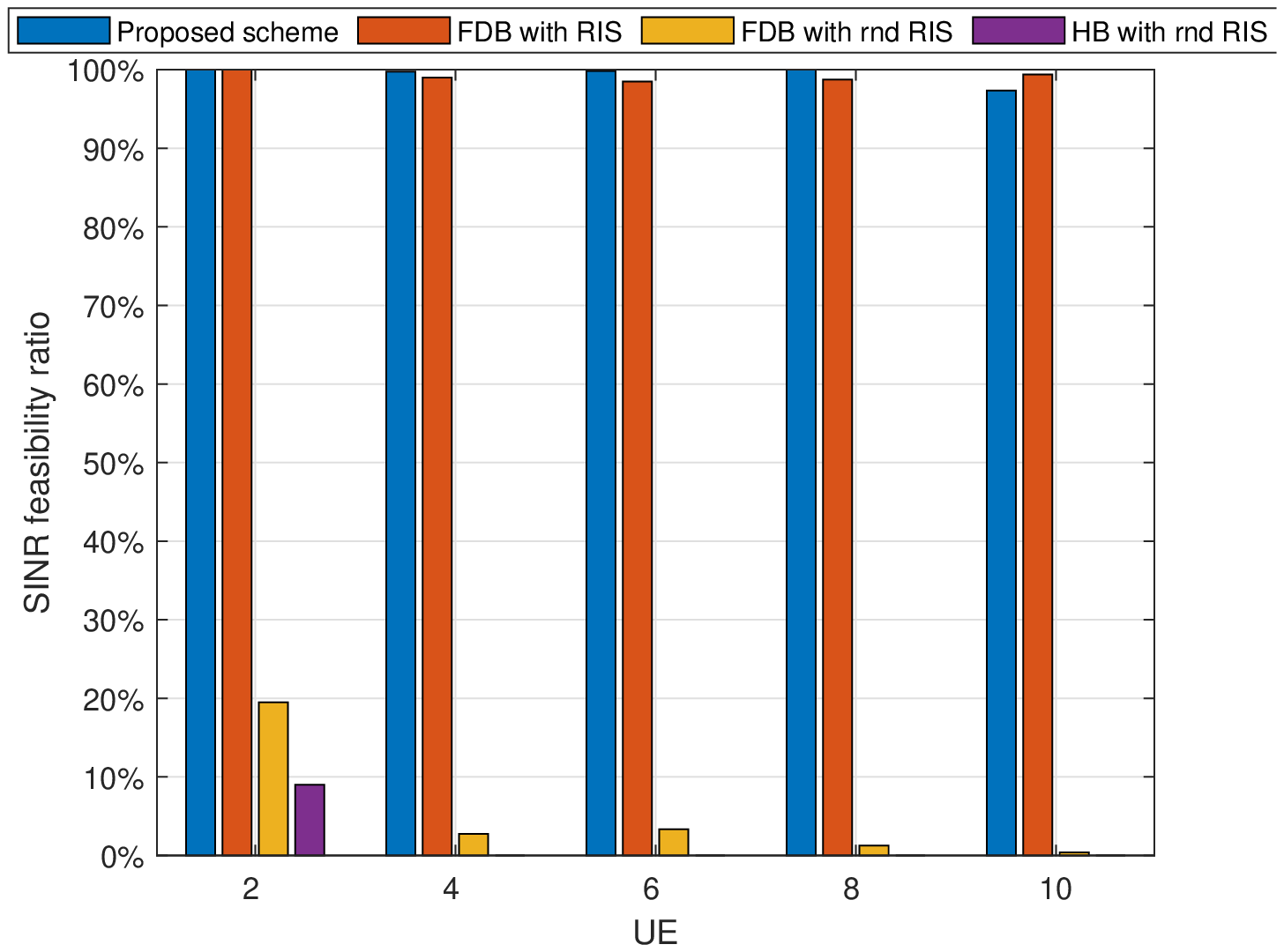}
		\end{minipage}
		\label{Result_UESINR}
	}
    	\subfigure[Beampattern MSE at the RIS]{
    		\begin{minipage}[b]{0.475\textwidth}
   		 	\includegraphics[width=1\textwidth]{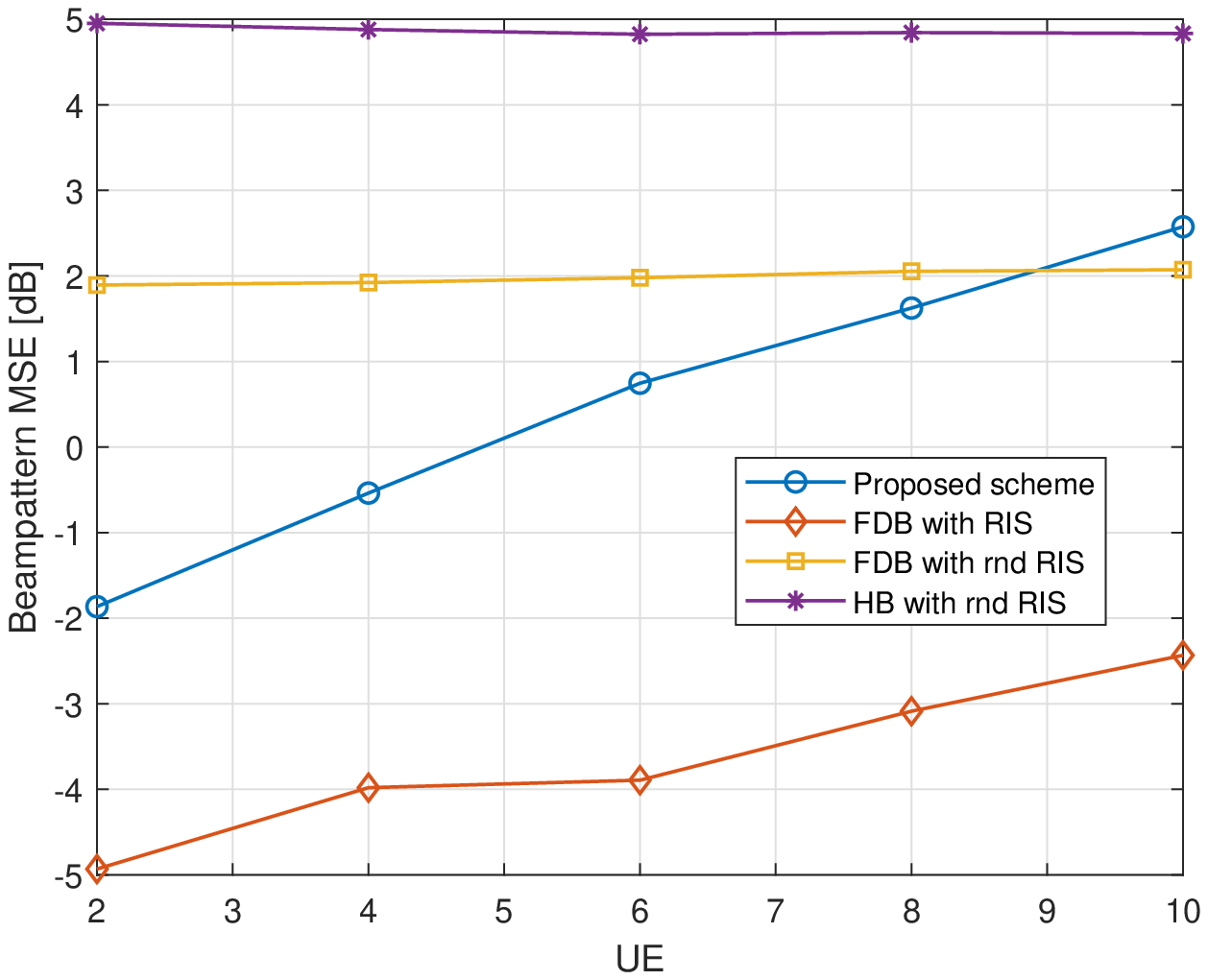}
    		\end{minipage}
		\label{Result_UEMSE}
    	}
	\caption{SINR feasibility ratio and beampattern MSE at the RIS under a varying number of UEs.}
	\label{Result_UE}
\end{figure}

Fig.~\ref{Result_UE} shows the SINR feasibility ratio and beampattern MSE at the RIS as a function of the number of UEs. In the proposed scheme and ‘FDB with RIS’ scheme, when the number of UEs increases, the SINR feasibility ratio almost maintains 100\% while the beampattern MSE worsens. This is because when the number of UEs increases, the multi-user interference on each sub-carrier increases as well. With the provision of the same transmit power, it becomes more difficult to satisfy the SINR constraints for all the UEs on all the sub-carriers. Therefore, more transmit power is allocated to UEs to compensate for the higher multi-user interference and thus to meet the SINR requirements. Consequently, less power remains available for sensing, thus leading to worse beampatterns. Besides, the SINR feasibility ratio and beampattern MSE of the proposed scheme are superior to those of ‘FDB with rnd RIS’ scheme and ‘HB with rnd RIS’ scheme. In particular, the proposed scheme has at least 80.5\% SINR feasibility improvement and 26.54\% beampattern MSE improvement compared to the two random RIS schemes, demonstrating that the existence of RIS improves the sensing and communication performances. Furthermore, the ‘FDB with RIS’ scheme has a slightly worse SINR feasibility ratio than the proposed scheme. This is because the alternating optimization framework leveraged in the ‘FDB with RIS’ scheme gets stuck in the local optimum.


\subsubsection{Impact of number of RIS elements}
\begin{figure}
	\centering
	\subfigure[SINR feasibility ratio]{
		\begin{minipage}[b]{0.475\textwidth}
			\includegraphics[width=1\textwidth]{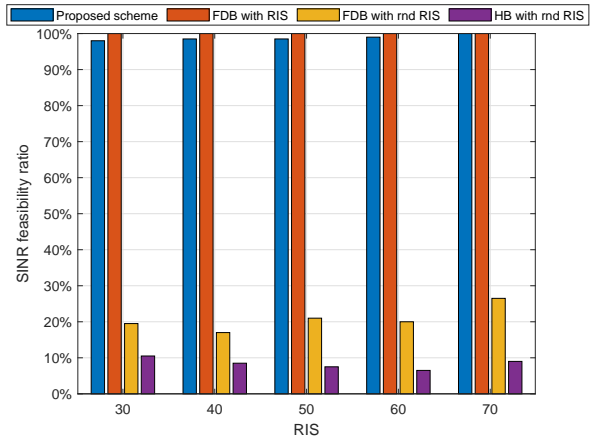}
		\end{minipage}
		\label{Result_RISSINR}
	}
    	\subfigure[Beampattern PSLR at the RIS at the RIS]{
    		\begin{minipage}[b]{0.475\textwidth}
   		 	\includegraphics[width=1\textwidth]{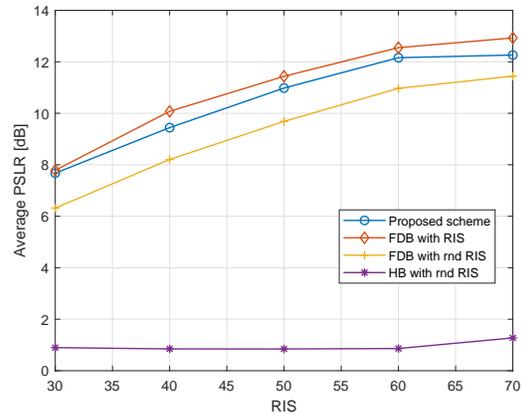}
    		\end{minipage}
		\label{Result_RISPSLR}
    	}
	\caption{SINR feasibility ratio and beampattern PSLR at the RIS under varying RIS sizes}
	\label{Result_RIS}
\end{figure}

To achieve better beamforming with more RIS elements, i.e., narrower beamwidth and higher gain, we generate the reference beampattern at the RIS based on the RIS size. In particular, more RIS elements result in a reference beampattern with narrower beamwidth and higher gain. Therefore, under different RIS sizes, the beampattern MSE obtained by different reference beampatterns can not be compared against each other. In this case, the metric average PSLR is employed for sensing performance comparison, measuring the ratio of the main lobe to the side lobe. Fig.~\ref{Result_RIS} illustrates the SINR feasibility ratio and beampattern PSLR versus the number of RIS elements. We observe that the average PSLRs in the proposed scheme and two fully-digital schemes increase as the RIS size increases. Besides, in our proposed scheme, the SINR feasibility ratio improves slightly when there are more RIS phase shifts. In contrast, the SINR feasibility ratio in schemes with random RIS phase shifts reduces by at least 74\% compared to the proposed scheme and does not show a regular pattern due to the random RIS phase configuration. Therefore, by including the RIS, the sensing and communication performances can be improved. In addition, more RIS elements result in higher performance improvement.

\subsubsection{Impact of SINR threshold}

\begin{figure}
	\centering
	\subfigure[SINR feasibility ratio]{
		\begin{minipage}[b]{0.475\textwidth}
			\includegraphics[width=1\textwidth]{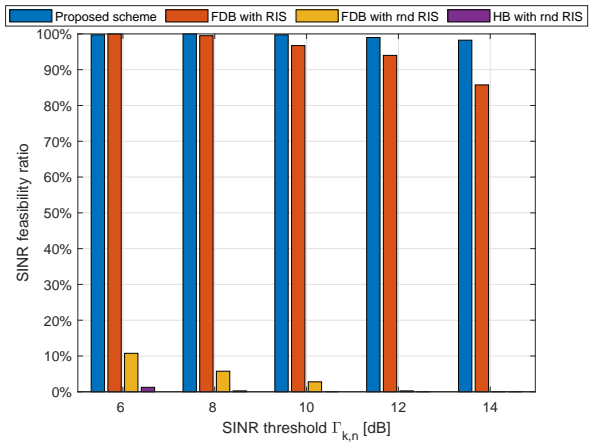}
		\end{minipage}
		\label{Result_GaSINR}
	}
    	\subfigure[Beampattern MSE at the RIS]{
    		\begin{minipage}[b]{0.475\textwidth}
   		 	\includegraphics[width=1\textwidth]{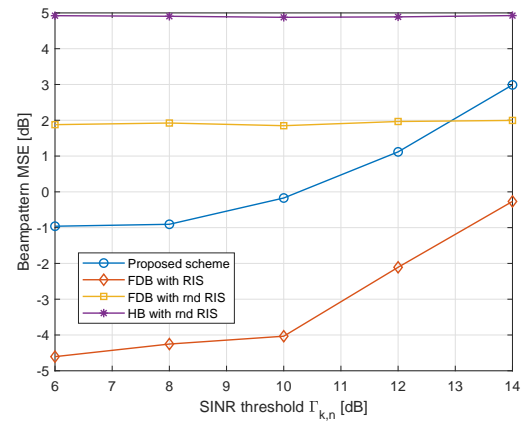}
    		\end{minipage}
		\label{Result_GaMSE}
    	}
	\caption{SINR feasibility ratio and beampattern MSE at the RIS under varying SINR requirements.}
	\label{Result_Ga}
\end{figure}

In Fig.~\ref{Result_Ga}, the influence of SINR threshold on the sensing and communication performances is shown. We observe that our proposed scheme and ‘FDB with RIS’ scheme can reach a feasible solution when the SINR threshold ${\Gamma _{n,k}}$ is below 12 dB. Beyond this threshold, we can use other measures, e.g., improving the transmit power or utilizing more RIS elements to reach a feasible solution. The SINR feasibility ratio of ‘FDB with RIS’ scheme is lower than the proposed scheme for the same reason as explained in Fig. {\ref{Result_UESINR}}, namely being stuck in the local optimum by the alternating optimization used in ‘FDB with RIS’ scheme. In contrast, the SINR performance of schemes with random RIS phase shifts is close to zero, because the RIS configuration does not contribute to improving communication. Besides, the beampattern MSEs of our proposed scheme and ‘FDB with RIS’ scheme increase as the SINR threshold grows. This is due to the fact that a higher transmit power is consumed to support the higher SINR requirements, leaving less transmit power for sensing.

\subsubsection{Impact of SNR}
\begin{figure}
	\centering
	\subfigure[SINR feasibility ratio]{
		\begin{minipage}[b]{0.475\textwidth}
			\includegraphics[width=1\textwidth]{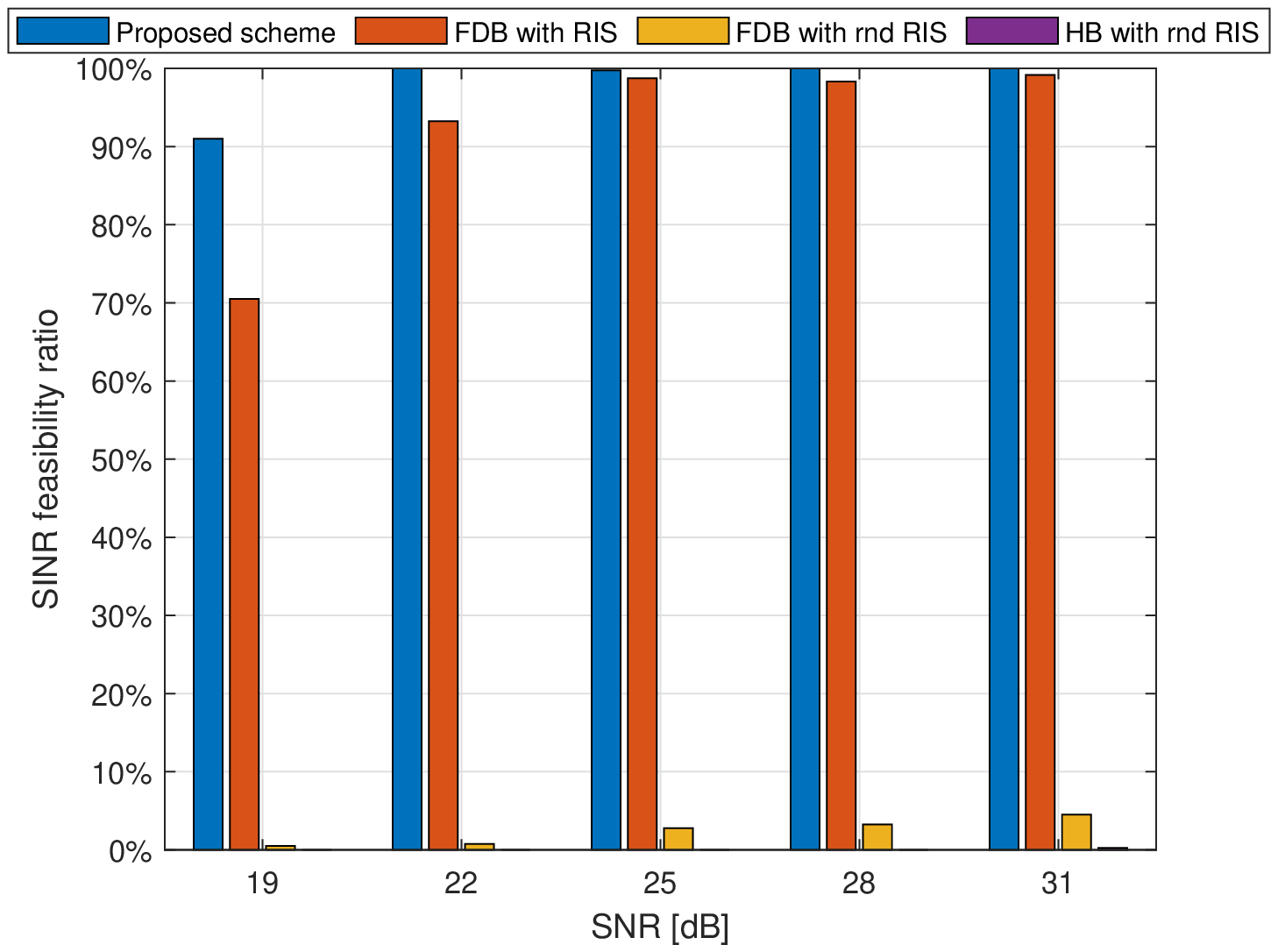}
		\end{minipage}
		\label{Result_SNRSINR}
	}
    	\subfigure[Beampattern MSE at the RIS]{
    		\begin{minipage}[b]{0.475\textwidth}
   		 	\includegraphics[width=1\textwidth]{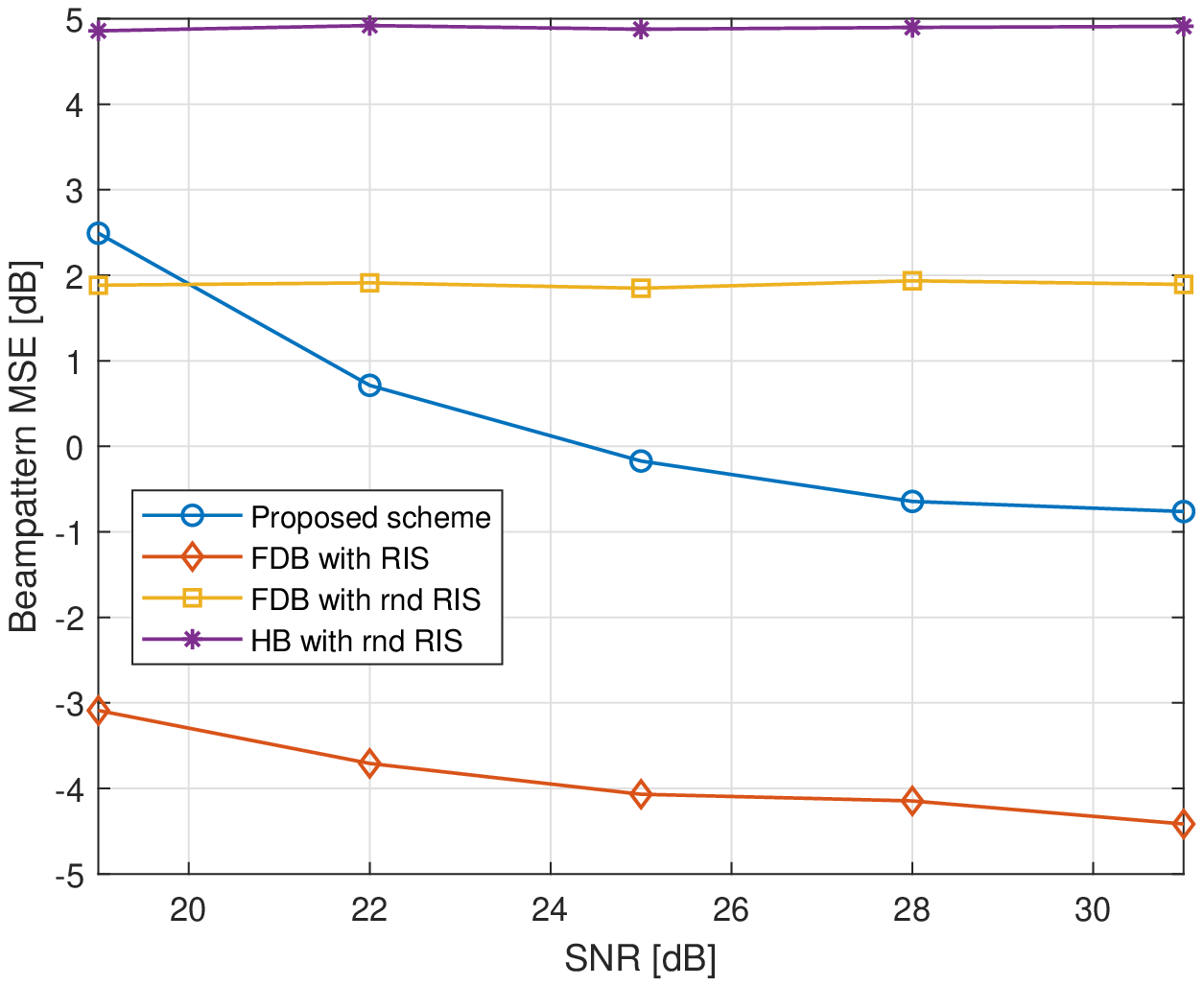}
    		\end{minipage}
		\label{Result_SNRMSE}
    	}
	\caption{SINR feasibility ratio and beampattern MSE at the RIS under varying}
	\label{Result_SNR}
\end{figure}

Fig.~\ref{Result_SNR} illustrates the impact of SNR on the sensing and communication performance. It can be observed that a higher SNR results in an increasing SINR feasibility ratio under the proposed scheme, ‘FDB with RIS’ scheme and ‘FDB with rnd RIS’ scheme. This is because, for a fixed transmit power, the noise power becomes smaller under a higher SNR, hence the SINR requirements are easier to be satisfied. Besides, the beampattern MSEs of our proposed scheme and ‘FDB with RIS’ scheme gradually decrease as the SNR increases. The reason is that the SINR can be satisfied more easily under a larger SNR, thus leaving more power to optimize sensing. 

In conclusion, in this section, we provide different parameter settings to evaluate their impact on both communication performance and sensing performance. From the simulation results, it can be observed that the sensing goal and communication SINR requirements can be satisfied by configuring suitable system parameters.

\subsection{Tradeoff between Communication and Sensing}

\begin{figure}
	\centering
	\subfigure[Beampattern MSE at the RIS versus SINR feasibility ratio]{
		\begin{minipage}[b]{0.475\textwidth}
			\includegraphics[width=1\textwidth]{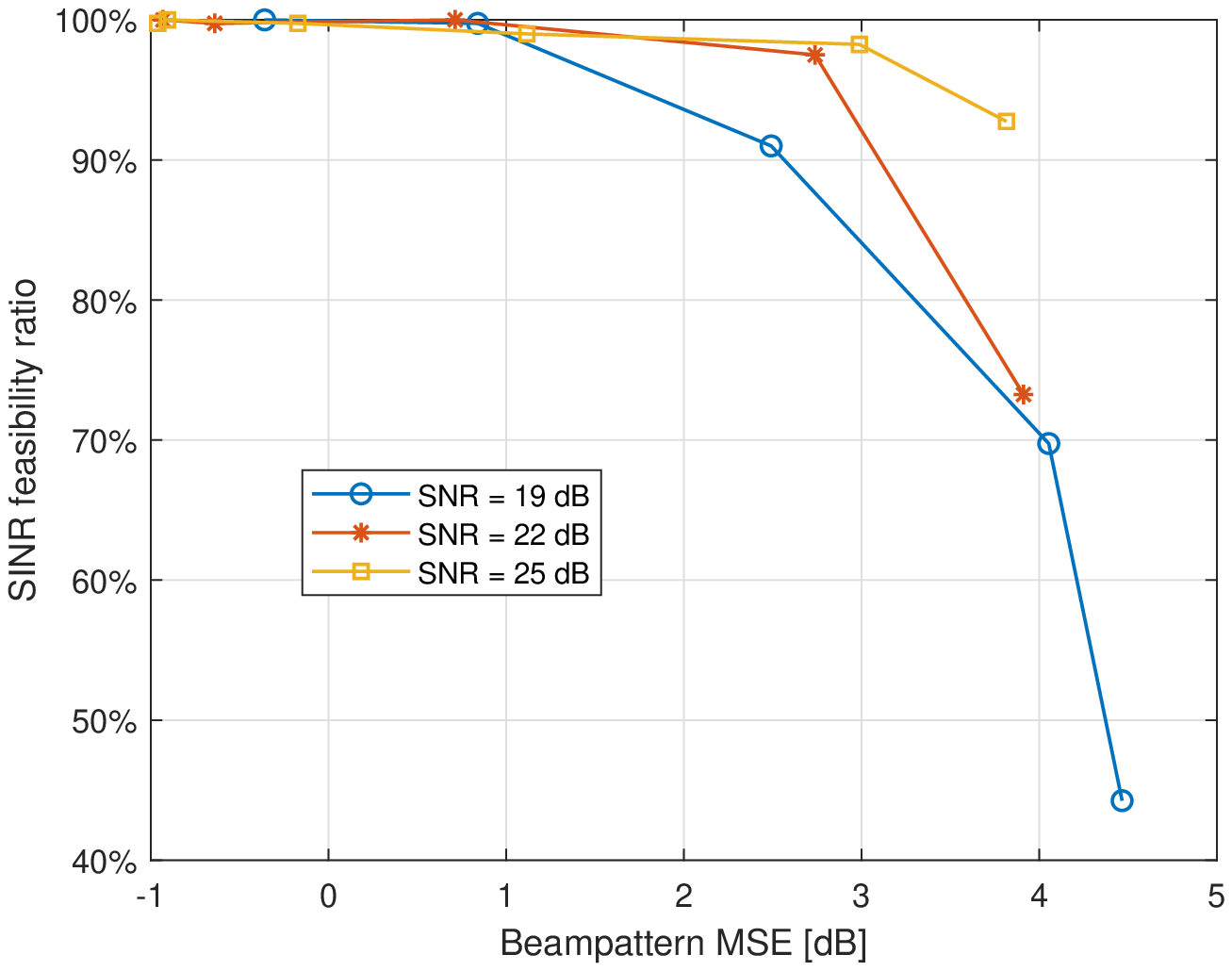}
		\end{minipage}
		\label{Tradeoff_SINRrvMSE}
	}
    	\subfigure[Beampattern MSE at the RIS versus Average SINR]{
    		\begin{minipage}[b]{0.475\textwidth}
   		 	\includegraphics[width=1\textwidth]{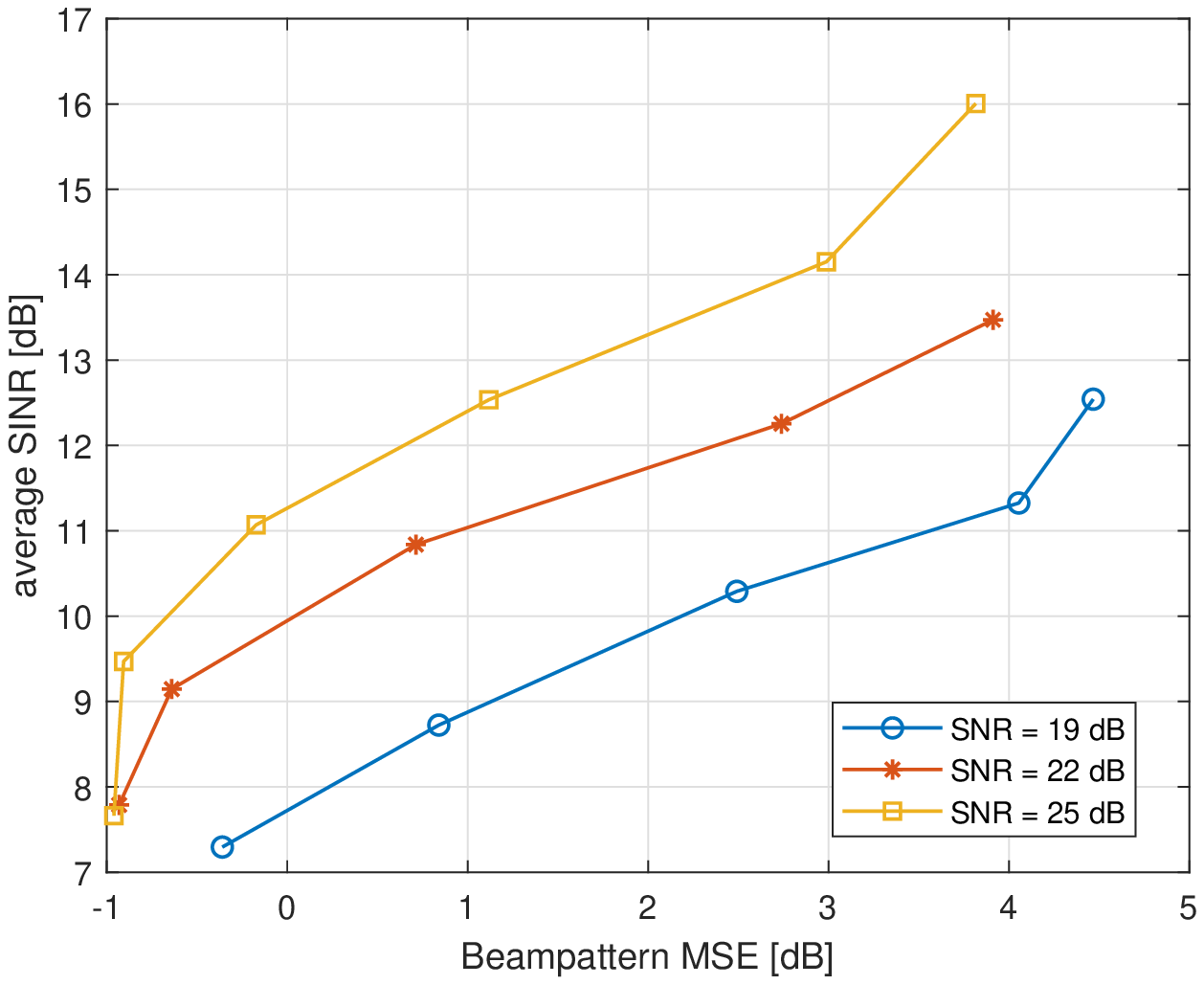}
    		\end{minipage}
		\label{Tradeoff_aveSINRvMSE}
    	}
	\caption{Tradeoff between communication performance and sensing performance}
	\label{Tradeoff_Ga}
\end{figure}

In this section, we measure the tradeoff between sensing performance and communication performance by analyzing the SINR feasibility ratio, the average SINR of all the UEs over all the sub-carriers, and the beampattern MSE, see Fig.~\ref{Tradeoff_Ga}. It is observed that reaching a higher beampattern MSE corresponds to a lower SINR feasibility ratio and a higher average SINR, i.e., improving communication performance has a negative impact on the sensing performance. For instance, in Fig.~\ref{Tradeoff_Ga},  we can notice that when SNR = 19 dB, reducing the beampattern MSE by 1.56 dB improves the SINR feasibility ratio by 21\% and decreases the average SINR by 1.04 dB. Furthermore, given a higher SNR, a higher average SINR or a lower beampattern MSE can be obtained. For example, in Fig.~\ref{Tradeoff_aveSINRvMSE}, when the SNR increases from 19 dB to 25 dB, up to 1.61 dB average SINR increment and 2.94 dB beampattern MSE increment are achieved. Hence, different tradeoffs between sensing performance and communication performance can be reached under different parameter configurations. For practical usage, the performances of communication and sensing can be weighed according to the required applications.



\section{Conclusion}
In this paper, we investigated the potential of RIS to improve the sensing and communication capabilities of the OFDM JCAS system at mmWave bands. To maximize the performance of sensing and communication, we formulated an optimization model via the joint hybrid beamforming and RIS phase shift design, in order to minimize the difference between the reference beampattern and the designed beampattern at the RIS subject to the power and SINR constraints. By introducing an auxiliary variable and leveraging the penalty method, a manifold-based ADMM algorithm was proposed to solve the formulated non-convex problem. Simulation results demonstrated that our proposed scheme can improve the SINR feasibility ratio for communication and the beampattern PSLR for sensing under larger RIS size. The tradeoff between sensing performance and communication performance was also discussed. As future work, we plan to model the mobility of targets and investigate its impact on the sensing performance.

\bibliographystyle{IEEEtran}
\bibliography{biblioJCAS}

\end{document}